%% file: main.tex
\documentclass[a4paper,11pt]{article}
\pdfoutput=1 
\usepackage{jcappub} 
\usepackage{xcolor,bm}
\usepackage{xspace}
\usepackage{hyperref}
\usepackage{graphicx}

\defcitealias{dmdb2020}{\texttt{dMdB20}}

\input{mydefinitions}

\title{3D Correlations in the Lyman-$\alpha$ Forest from Early DESI Data}  

\author[1]{C. Gordon,}
\author[2]{A. Cuceu\footnote{NASA Einstein Fellow},}
\author[1]{J. Chaves-Montero,}
\author[1,3]{A. Font-Ribera,}
\author[4,5]{A. X. González-Morales,}

\author[6]{J.~Aguilar,}
\author[7]{S.~Ahlen,}
\author[8]{E.~Armengaud,}
\author[6]{S.~Bailey,}
\author[9]{A.~Bault,}
\author[10]{A.~Brodzeller,}
\author[3]{D.~Brooks,}
\author[6]{T.~Claybaugh,}
\author[5]{R.~de la Cruz,}
\author[10]{K.~Dawson,}
\author[3]{P.~Doel,}
\author[11,12]{J.~E.~Forero-Romero,}
\author[6]{S.~Gontcho A Gontcho,}
\author[6]{J.~Guy,}
\author[5]{H.~K.~Herrera-Alcantar,}
\author[13]{V.~Ir\v{s}i\v{c},}
\author[2]{N.~G.~Kara{\c c}ayl{\i},}
\author[9]{D.~Kirkby,}
\author[6]{M.~Landriau,}
\author[14]{L.~Le~Guillou,}
\author[6]{M.~E.~Levi,}
\author[15]{A.~de la Macorra}

\author[1,16]{M.~Manera,}

\author[2]{P.~Martini,}
\author[17]{A.~Meisner,}
\author[1,18]{R.~Miquel,}
\author[19]{P.~Montero-Camacho,}
\author[15]{A.~Muñoz-Gutiérrez,}
\author[20]{L.~Napolitano,}
\author[21]{J.~Nie,}
\author[5,22]{G.~Niz,}
\author[6,8]{N.~Palanque-Delabrouille,}
\author[23,24]{W.~J.~Percival,}
\author[25]{M.~Pieri,}
\author[6,26]{C.~Poppett,}
\author[27]{F.~Prada,}
\author[28]{I.~P\'erez-R\`afols,}
\author[1]{C.~Ram\'irez-P\'erez,}
\author[8,29]{C.~Ravoux,}
\author[30]{M.~Rezaie,}
\author[2]{A.~J.~Ross,}
\author[31]{G.~Rossi,}
\author[32]{E.~Sanchez,}
\author[6]{D.~Schlegel,}
\author[33]{M.~Schubnell,}
\author[34]{H.~Seo,}
\author[35]{F.~Sinigaglia,}
\author[14]{T.~Tan,}
\author[36]{G.~Tarl\'{e},}
\author[37]{M.~Walther,}
\author[16]{B.~A.~Weaver,}
\author[8]{C.~Yèche,}
\author[21]{Z.~Zhou,}
\author[21]{H.~Zou}

\affiliation[1]{Institut de F\'{i}sica d’Altes Energies(IFAE),The Barcelona Institute of Science and Technology, 08193 Bellaterra (Barcelona), Spain}
\affiliation[2]{Center for Cosmology and Astro-Particle Physics, The Ohio State University, Columbus, Ohio 43210, USA}
\affiliation[3]{Department of Physics and Astronomy, University College London, Gower Street, London WC1E 6BT, UK}
\affiliation[4]{Consejo Nacional de Ciencia y Tecnología, Av. Insurgentes Sur 1582. Colonia Crédito Constructor, Del. Benito Juárez C.P.03940, México D.F. México}
\affiliation[5]{Departamento de Física, Universidad de Guanajuato - DCI, C.P. 37150, Leon, Guanajuato, México}
\affiliation[6]{Lawrence Berkeley National Laboratory, 1 Cyclotron Road, Berkeley, CA 94720, USA}
\affiliation[7]{Physics Dept., Boston University, 590 Commonwealth Avenue, Boston, MA 02215, USA}
\affiliation[8]{IRFU, CEA, Universit\'{e} Paris-Saclay, F-91191 Gif-sur-Yvette, France}
\affiliation[9]{Department of Physics and Astronomy, University of California, Irvine, 92697, USA}
\affiliation[10]{Department of Physics and Astronomy, The University of Utah, 115 South 1400 East, Salt Lake City, UT 84112, USA}
\affiliation[11]{Observatorio Astron\'omico, Universidad de los Andes, Cra. 1 No. 18A-10, Edificio H, CP 111711 Bogot\'a, Colombia}
\affiliation[12]{Departamento de F\'isica, Universidad de los Andes, Cra. 1 No. 18A-10, Edificio Ip, CP 111711, Bogot\'a, Colombia}
\affiliation[13]{Kavli Institute for Cosmology, University of Cambridge, Madingley Road, Cambridge CB3 0HA, UK}
\affiliation[14]{Sorbonne Universit\'{e}, CNRS/IN2P3, Laboratoire de Physique Nucl\'{e}aire et de Hautes Energies (LPNHE), FR-75005 Paris, France}
\affiliation[15]{Instituto de F\'{\i}sica, Universidad Nacional Aut\'{o}noma de M\'{e}xico,  Cd. de M\'{e}xico  C.P. 04510,  M\'{e}xico}

\affiliation[16]{Departament de F\'{i}sica, Serra H\'{u}nter, Universitat Aut\`{o}noma de Barcelona, 08193 Bellaterra (Barcelona), Spain}
\affiliation[17]{NSF's NOIRLab, 950 N. Cherry Ave., Tucson, AZ 85719, USA}
\affiliation[18]{Instituci\'{o} Catalana de Recerca i Estudis Avan\c{c}ats, Passeig de Llu\'{\i}s Companys, 23, 08010 Barcelona, Spain}
\affiliation[19]{Department of Astronomy, Tsinghua University, 30 Shuangqing Road, Haidian District, Beijing, China, 100190}
\affiliation[20]{Department of Physics \& Astronomy, University  of Wyoming, 1000 E. University, Dept.~3905, Laramie, WY 82071, USA}
\affiliation[21]{National Astronomical Observatories, Chinese Academy of Sciences, A20 Datun Rd., Chaoyang District, Beijing, 100012, P.R. China}
\affiliation[22]{Instituto Avanzado de Cosmolog\'{\i}a A.~C., San Marcos 11 - Atenas 202. Magdalena Contreras, 10720. Ciudad de M\'{e}xico, M\'{e}xico}
\affiliation[23]{Perimeter Institute for Theoretical Physics, 31 Caroline St. North, Waterloo, ON N2L 2Y5, Canada}
\affiliation[24]{Waterloo Centre for Astrophysics, University of Waterloo, 200 University Ave W, Waterloo, ON N2L 3G1}
\affiliation[25]{Aix Marseille Univ, CNRS, CNES, LAM, Marseille, France}
\affiliation[26]{Space Sciences Laboratory, University of California, Berkeley, 7 Gauss Way, Berkeley, CA  94720, USA}
\affiliation[27]{Instituto de Astrof\'{i}sica de Andaluc\'{i}a (CSIC), Glorieta de la Astronom\'{i}a, s/n, E-18008 Granada, Spain}
\affiliation[28]{Departament de F\'{\i}sica Qu\`{a}ntica i Astrof\'{\i}sica, Universitat de Barcelona, Mart\'{\i} i Franqu\`{e}s 1, E08028 Barcelona, Spain}
\affiliation[29]{Aix Marseille Univ, CNRS/IN2P3, CPPM, Marseille, France}
\affiliation[30]{Department of Physics, Kansas State University, 116 Cardwell Hall, Manhattan, KS 66506, USA}
\affiliation[31]{Department of Physics and Astronomy, Sejong University, Seoul, 143-747, Korea}
\affiliation[32]{CIEMAT, Avenida Complutense 40, E-28040 Madrid, Spain}
\affiliation[33]{Department of Physics, University of Michigan, Ann Arbor, MI 48109, USA}
\affiliation[34]{Department of Physics \& Astronomy, Ohio University, Athens, OH 45701, USA}
\affiliation[35]{Instituto de Astrof\'{i}sica de Canarias, C/ Vía L\'{a}ctea, s/n, E-38205 La Laguna, Tenerife, Spain}
\affiliation[36]{University of Michigan, Ann Arbor, MI 48109, USA}
\affiliation[37]{University Observatory, Faculty of Physics, Ludwig-Maximilians-Universit\"{a}t, Scheinerstr. 1, 81677 M\"{u}nchen, Germany}

\emailAdd{cgordon@ifae.es}
\input{abstract.tex}

\begin{document}

\maketitle

\flushbottom

\input{introduction.tex}
\input{data.tex}
\input{correlations.tex}

\input{model.tex}
\input{fitting.tex}
\input{conclusion.tex}

\appendix
\input{corner_plots.tex}

\acknowledgments
\input{acknowledgments.tex}

\input{zenodo.tex}

\bibliographystyle{JHEP.bst}
\bibliography{biblio}

\end{document}

%% file: mydefinitions.tex
\newcommand{\lya}{Ly$\alpha$\xspace}

\newcommand{\lyaf}{Ly$\alpha$ forest\xspace}

\newcommand{\dMdB}{\citetalias{dmdb2020}}

%% file: abstract.tex
\abstract{{We present the first measurements of Lyman-$\alpha$ (\lya) forest correlations using early data from the Dark Energy Spectroscopic Instrument (DESI)}. {We measure the auto-correlation of \lya absorption using 88\,509 quasars at $z>2$, and its cross-correlation with quasars using a further 147\,899 tracer quasars at $z\gtrsim1.77$.} Then, we fit these correlations using a 13-parameter model based on linear perturbation theory and find that it provides a good description of the data across a broad range of scales. {We detect the BAO peak with a signal-to-noise ratio of $3.8\sigma$, and show that our measurements of the auto- and cross-correlations are fully-consistent with previous measurements by the Extended Baryon Oscillation Spectroscopic Survey (eBOSS).}
{Even though we only use here a small fraction of the final DESI dataset, {our uncertainties} are only a factor of 1.7 larger than those from the final eBOSS measurement}. {We validate the existing analysis methods of \lya correlations in preparation for making a robust measurement of the BAO scale with the first year of DESI data.}}

%% file: introduction.tex
\section{Introduction}\label{sec:int}
{The Lyman-$\alpha$ (Ly$\alpha$) forest is a dense series of absorption lines found in the spectra of distant quasars, that we use to study the large-scale structure of the universe}. These absorption features arise as a result of diffuse neutral hydrogen in the intergalactic galactic medium \cite{Lynds}, which readily absorbs light at the Ly$\alpha$ wavelength $\lambda_{{\rm Ly}\alpha}=1215.67\mathrm{\AA}$, and can be used as a biased tracer of the underlying matter density \cite{Weinberg_2003}. Due to the low rest-frame wavelength of the Ly$\alpha$ transition, its absorption is only {detected by optical spectrographs for} quasars with redshift $z\geq2$. {The \lyaf allows us to measure the expansion of the Universe at redshifts larger than those accessible by spectroscopic galaxy surveys ($z<$1.5, \cite{eboss_2021}).}

The environment that gives rise to the Ly$\alpha$ forest has been well studied using hydro-dynamical simulations \cite{Cen94_sim_lya,Escude96}, where a combination of photo-ionisation heating and adiabatic cooling leads to a tight relation between temperature and gas density $T\propto(\rho/\overline{\rho})^{\gamma-1}$ \cite{Hui_1997}. 
The transmitted flux fraction is related to the optical depth of the Ly$\alpha$ forest by $F=e^{-\tau}$. 
At sufficiently large scales ($>10h^{-1}$Mpc), the power spectrum of transmitted flux  $\delta_\mathrm{F} = {F}/{\overline{F}} -1$, is found to trace {linearly} the power spectrum of $\delta(\rm x) = \rho(\mathrm{x})/\overline{\rho} - 1$ \cite{Slosar_2011}. 
However, due to non-linear growth, thermal broadening, and Jeans smoothing, this relation becomes more complex and non-linear at smaller scales \cite{McDonald2003,Arinyo_i_Prats_2015,Givans_2022}. 

{The first studies of clustering in the \lyaf looked at one-dimensional (1D) correlations along the line-of-sight towards a handful of quasars \cite{croft98,mcdonald2000}.}
{These 1D measurements were later repeated using larger data sets from the Sloan Digital Sky Survey (SDSS; \cite{McDonald2006}), from the Baryon Acoustic Oscillation Survey (BOSS; \cite{Palanque_Delabrouille_2013}) and from its extension (eBOSS; \cite{Chabanier_2019}).}

{This paper, on the other hand, focuses on three-dimensional (3D) correlations in the \lyaf.
Using early data from the BOSS survey, \cite{Slosar_2011} presented the first measurement of 3D correlations in the \lyaf.
These correlations were also computed from larger datasets to measure the scale of Baryon Acoustic Oscillations (BAO) around $z=2.3$, in BOSS and eBOSS \cite{Busca2013,slosar2013,Kirkby2013,Delubac2015,B17,dSA2019,dmdb2020}. 
The cross-correlation of quasars and the \lyaf was also measured in BOSS \cite{FontRibera2013}, and it was used to improve the BAO measurements from the \lya datasets \cite{Font_Ribera_2014,dmdb2017,Blomqvist2019,dmdb2020}. 
{The final eBOSS analysis \cite{dmdb2020}, using data from SDSS DR16, included 210\,005 \lya forest from quasars at $z>$2.1 that were used in the auto-correlation, and 341\,468 at $z>$1.77 that were also used in the cross-correlation}. 
More recently, the full shape of the 3D Ly$\alpha$ correlation has been used to measure the Alcock-Paczynski effect in eBOSS data \cite{Cuceu2022_data}, which combined with Ly$\alpha$ BAO gives the best measurement of the expansion rate of the Universe above $z=1$.}

{We use early data from the Dark Energy Spectroscopic Instrument (DESI) \cite{snowmass2013,desi_pt1} to measure the auto-correlation of the \lyaf and its cross-correlation with quasars, and compare our results with those from the eBOSS collaboration \cite{dmdb2020}.} {We also show in section \ref{sec:fitter} that we detect the BAO peak to high-confidence in our dataset. Given the early stage and relatively low statistical power of our data, we choose not to present measurements of the BAO scale parameters.}
Throughout the latter half of the paper, we make use of the best fit $\Lambda$CDM cosmological parameters from Planck 2018 \cite{Planck2018}\footnote{See table 2 
 of \cite{Planck2018}, where we use \textit{TT,TE,EE+lowE+lensing} results.}, where $\Omega_{\rm M}=0.3153$, to compute correlations in terms of co-moving separation. This allows us to combine a wide range of redshifts into a single measurement while preserving the BAO feature.

The structure of the paper follows roughly the steps required to perform the end-to-end analysis. We start by describing the DESI instrument and the early data release, and outline how we measure the Ly$\alpha$ transmission catalogue from high-redshift quasar spectra in section \ref{sec:data}.
We then compute the auto- and cross-correlation and covariance matrices in section \ref{sec:correlations}. 
In section \ref{sec:modelling}, we discuss the physical model of \lyaf correlations and contaminants, and in section \ref{sec:fitter} we present the results of fitting this model to our data and discuss the consistency of these results with the final eBOSS analysis \cite{dmdb2020}. 
Finally, we summarise our results in section \ref{sec:conclusion}.

{The \lyaf dataset used here is described in detail in \cite{cper}. Our analysis is also accompanied by a paper analysis on early DESI mock datasets \cite{hiram} which utilises the same 13-parameter model to describe the auto- and cross-correlation. We further highlight two papers where systematic effects, including quasar redshift errors \cite{abault} and instrumental effects \cite{Satya}, are studied on the same early DESI data set.}

%% file: data.tex
\section{Data}\label{sec:data}

{DESI is a multi-object spectrograph installed at the Mayall 4-meter telescope at Kitt Peak National Observatory \cite{Instrument_Overview_KP}. DESI is equipped with 5000 robotically controlled fibres across 10 independent spectrographs, that have an approximately 3-degree field of view in each pointing \cite{desi_pt2,Focal_Plane,Corrector}.}
{Each measured spectrum covers an observed wavelength range between 3600 and 9800$\mathrm{\AA}$ with a pixel width of 0.8$\mathrm{\AA}$, which is split between 3 independent spectrograph arms ranging between 3600 -- 5930, 5600 -- 7720, and 7470 -- 9800$\mathrm{\AA}$ respectively \cite{Instrument_Overview_KP}.}

{In May 2021, DESI started a 5-year campaign which will observe 3 million quasar \cite{Chaussidon_2023} and 36 million galaxy targets \cite{HahnBGS,ZhouLRG,RaichoorELG}, over an area of 14\,000 deg$^2$.
These targets are selected from the DESI Legacy Imaging Survey \cite{Dey_2019,BASS,dr9}.} 
{These observations are made thanks to extensive supporting software packages: one to select targets for spectroscopic observation \cite{Myers_2023}, one to assign fibres to targets \cite{fiberassign}, one to optimise survey operations \cite{ops}, an exposure-time calculator \cite{exp_time_calc}, and a spectroscopic pipeline to reduce the data and calibrate the spectra \cite{JGuy}.}

We analyse data from the Early Data Release (EDR) of DESI \cite{EDR}, which is publicly available{\footnote{{https://data.desi.lbl.gov/public/edr/}}} and contains 68\,750 quasars. 
EDR contains data from a pre-survey designed to optimise to optimise target selection algorithms \cite{SVoverview,Chaussidon_2023}, refine the observational procedure, and generally evaluate the quality of spectra and the redshift accuracy \cite{qso_VI,qso_templates}.
To increase the statistical power of the measurements, we also include data from the first two months of the main survey (M2) containing 249\,941 quasars. We use "EDR+M2" to refer to the combination of both data sets. 

\begin{figure}
    \centering
    \includegraphics[width=1\textwidth]{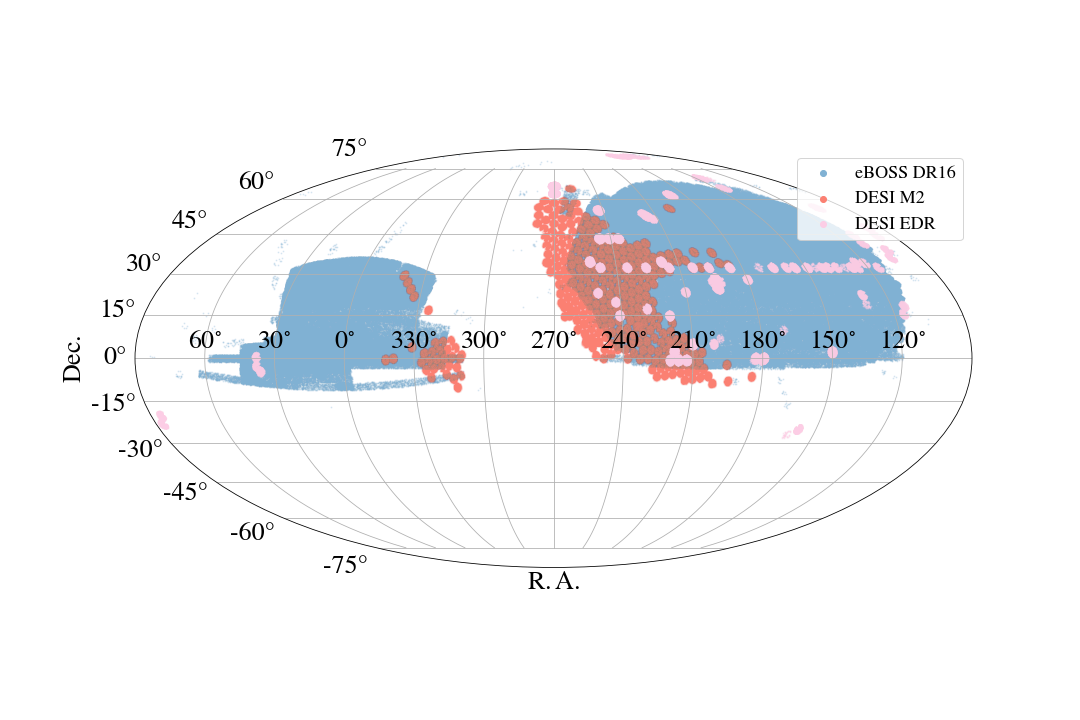}
    \caption{Footprint of quasar targets in DESI Early Data Release (purple), the first two months of main survey (red), and eBOSS DR16 (blue) \cite{dmdb2020}. The EDR and M2 surveys cover an area of 250.1 deg$^2$ and 1290.9 deg$^2$.}
    \label{fig:qso_footprint}
\end{figure}

The sample footprint is shown in figure \ref{fig:qso_footprint}, where we highlight the each survey phase and the eBOSS Data Release 16 (DR16) survey footprint \cite{dmdb2020}. 

Even though we analyse all sources from EDR+M2 together, the typical signal-to-noise (SNR) of M2 and EDR spectra varies. Targets observed in M2 are limited to one observing pass{\footnote{{Equivalent to nominal 1000s of effective exposure time, which is exposure time accounting for observing conditions and fibre effects.}}}, which will increase to 4 by the end of DESI observations. Quasars identified as Ly$\alpha$ quasars ($z>$2) receive more passes in general than other targets to reach the required SNR per pixel across the Ly$\alpha$ forest. On the other hand, roughly half of EDR spectra present 4 passes and half 1.3 \cite{Chaussidon_2023}.

\subsection{Quasar spectra}

The DESI pipeline uses a robust spectroscopic reduction pipeline \cite{JGuy}, in combination with a template-fitting algorithm \texttt{Redrock} (RR) \cite{rrock} to give object classifications and estimate their redshifts. Templates used by RR are formed by a linear combination of bases resulting from the principal component analysis (PCA) decomposition of Sloan Digital Sky Survey spectra \cite{rrock} {(see \cite{qso_templates} for comparison when using a new set of quasar templates)}.

To quantify the ability of the pipeline to classify quasars we use two terms, purity and completeness. Purity is the percentage of classified  quasars that are true quasars, which is $>$99\% in our pipeline. Completeness is the percentage of true quasars identified as quasars, which we find to be $\sim$86\%. To increase this, we run two algorithms on the RR output. The first attempts to identify the {Mg}{II} broad emission line -- a feature only found in quasars -- in the spectra of sources classified as galaxies \cite{Chaussidon_2023}, and is especially effective at 0.5$<z<$1.5. The other is \texttt{QuasarNet} (QN) \cite{Busca_2018,Farr_2020}, a convolutional neural network trained using a large number of visually inspected BOSS spectra, to identify faint quasars missed by both RR and the {first afterburner}, and is more useful at  $z\gtrsim2$. These algorithms increase the spectral completeness from $\sim$86 to $\sim$94\% \cite{qso_VI}, and only cause a $\sim$1-2\% drop in spectral purity at $z<1$ \cite{Chaussidon_2023}. 

\texttt{Redrock} is very reliable in estimating the redshifts of high-quality spectra \cite{qso_VI}.{The number of catastrophic redshifts, where the difference between visually-inspected and RR redshifts is $\Delta z>0.01$, is only $\sim$1.5\%}. For sources presenting a disagreement between RR and QN redshifts, the pipeline computes the final redshift by re-running RR using the QN redshift as prior and only quasar templates. This reduces the fraction of catastrophic redshifts from 1.5 to 1\% \cite{qso_VI}.

\begin{figure}
\hspace*{0cm}
    \centering
    \includegraphics[width=1\textwidth]{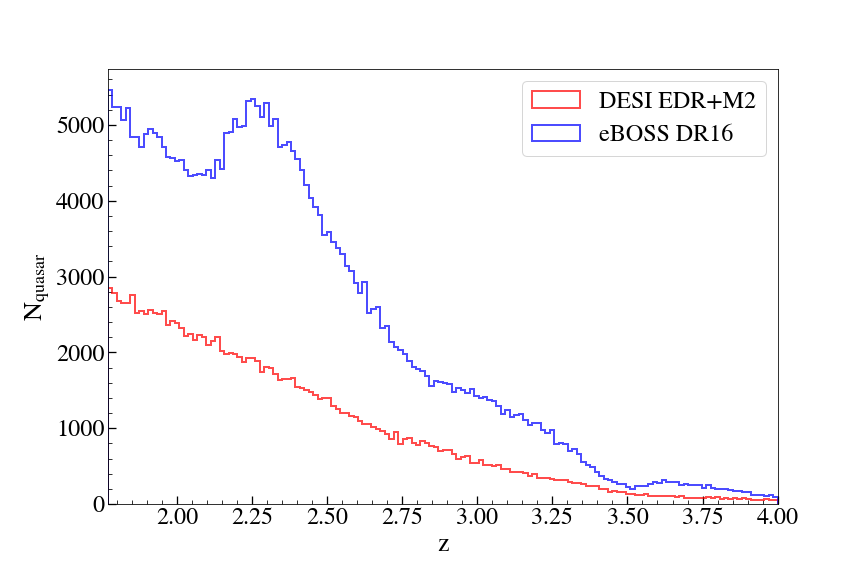}
    \caption{Redshift distribution of high-redshift quasars in DESI EDR+M2 and eBOSS DR16 \cite{dmdb2020}, with 147\,899 and 341\,468 in each sample respectively. Only quasars between $z=1.77$ and $z=3.75$ feature in our analysis.} 
    \label{fig:qso_pix_distrib}
\end{figure}

We obtain an overall density of 210 quasars deg$^{-2}$ (60 for $z>$2.1) \cite{Chaussidon_2023}, which surpasses the DESI requirements of 170 quasars deg$^{-2}$ (50 at $z>$2.1) \cite{desi_pt1}.
Throughout this work, we will only analyse spectroscopically identified quasars that were also selected as quasar targets (i.e. we ignore the small fraction of quasars found in other target classes) \cite{Myers_2023}, to keep the purity of the catalogue as high as possible. {These have a density of around 54 deg$^{-2}$ at $z>$2.1}. In figure \ref{fig:qso_pix_distrib} we show the redshift distributions of our EDR+M2 \lya quasars and include for comparison the catalogue used in the Ly$\alpha$ analysis of eBOSS DR16 (\cite{dmdb2020}, hereby {\dMdB}).

\begin{figure}
    \centering
    \includegraphics[width=0.9\textwidth]{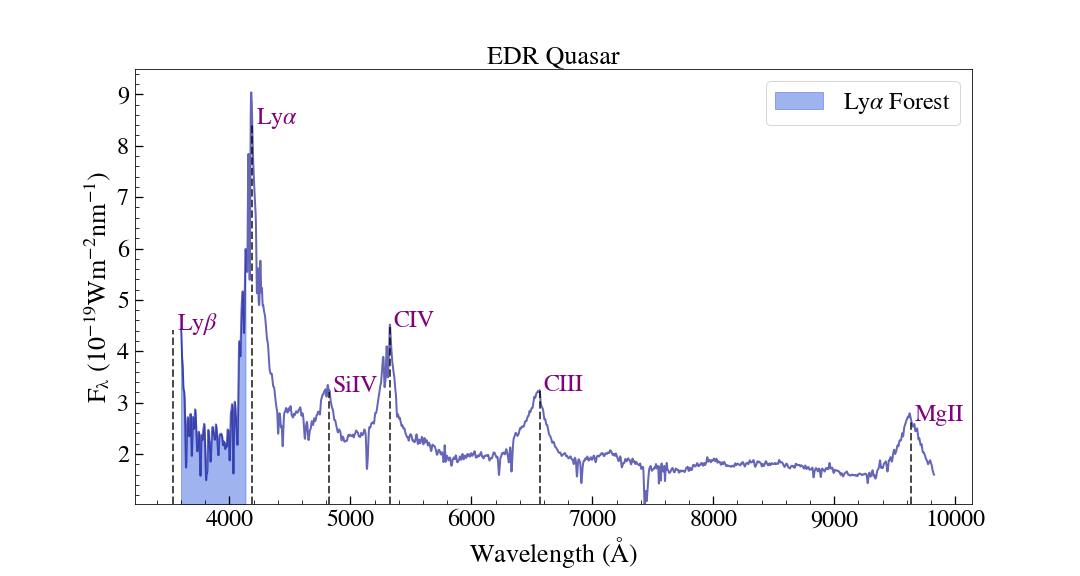}
    \caption{An example of a DESI EDR spectra at $z=2.44$. Black dashed lines indicate the central wavelength of the strongest quasar emission lines, including the Ly$\alpha$ used in our analysis. The shaded blue region shows the Ly$\alpha$ forest between the Ly$\alpha$ and Ly$\beta$ lines, the latter only partially observed in this specific example. Also of note is the MgII line, used in the pipeline to re-classify galaxies as quasars and the region between the CIV and CIII] lines, used for the re-calibration step described in section \ref{sec:lyacat}.} 
    \label{fig:spectra}
\end{figure}

Figure \ref{fig:spectra} shows one such quasar with $z=2.44$. Purple labels indicate broad, distinctive emission lines often observed in high-redshift quasars, including the MgII line. We also highlight the Ly$\alpha$ forest in blue, which exhibits a high density of absorption lines. Due to the rest-frame wavelength range of the \lyaf and the observed redshift distribution of quasars, most of our data falls within the blue arm of the spectrographs.

\subsection{Pixel masking}\label{sec:pixmask}

{The \lya catalogue used to obtain the results in this paper is described in \cite{cper}.
We present here a summary of how we mask contaminants in the data, and refer the reader to \cite{cper} for a detailed description of the process.}

In this section and while computing the flux transmission field, we use Ly$\alpha$ pixels of width 0.8$\mathrm{\AA}$ {in the observed frame} - the resolution provided by the spectroscopic pipeline \cite{JGuy}. When computing the correlation functions (section \ref{sec:correlations}) and the metal matrices (section \ref{sec:metals}), we combine these into 2.4$\mathrm{\AA}$ analysis pixels to reduce the computation time, without affecting the measured correlations.

The two primary contaminants of our \lya forests are broad absorption line features (BALs) and damped \lya absorbers (DLAs). BALs are broad spectral troughs that arise from high-velocity gas outflows from quasars. They often fall into the Ly$\alpha$ forest region and are difficult to model. These have been identified in the past using convolutional neural networks \cite{Guo_2019}, {and a similar approach has been used in DESI \cite{sfil}.}
Identified BALs are then masked in the affected spectra depending on the width of the trough, {following the procedure described in \cite{Ennesser_2022}.}
Masking avoids discarding whole Ly$\alpha$ forests where there are often valuable pixels, as was done in previous \lyaf BAO studies, like {\dMdB}.

DLAs are large absorption systems with column densities of neutral hydrogen $N_{\rm HI}>10^{\mathrm{20.3}}{\rm cm}^{-\mathrm{2}}$. Like the \lyaf, DLAs are a tracer of the underlying matter distribution but present a higher bias than the Ly$\alpha$ forest \cite{FontRibera2012,PerezRafols_2018,PerezRafols22}. 
Their wide absorption profiles, however, increase the noise in the clustering measurements. 
Following {\dMdB}, we mask the regions where more than 20\% of flux is absorbed DLAs and correct their damping wings in the unmasked regions using a Voigt profile.
We use a DLA catalog created using two different techniques, a convolutional neural network (CNN) and a Gaussian process (GP) finder, used in previous \lya analyses \cite{Parks_2018, Ho_2021} and re-developed for DESI \cite{Wang_2022,jzhou}.
{We decide to mask only those DLAs that are detected by both algorithms with a confidence higher than 50\%, resulting in 7\% (or $\sim$8500) of the forests in our analysis containing at least one DLA.}

To further remove noise from our data, we perform two additional steps. {The first is to mask areas affected by sharp spectral features, like the bright [OI] $5577.3\rm\AA$ sky line and the calcium H and {calcium} K lines from the interstellar medium. We mask the wavelength ranges between [5570.5,5586.7]$\mathrm{\AA}$, [3967.3,3971.0]$\mathrm{\AA}$ and [3933.0,3935.8]$\mathrm{\AA}$ respectively to avoid these\cite{cper}.} Other features caused by systematic flux calibration errors are corrected using the CIII] spectral region (between CIV and CIII] in figure \ref{fig:spectra}), see \cite{cper} for more details.

\subsection{The flux transmission field}\label{sec:lyacat}

Here we define the limits on our Ly$\alpha$ forests, then describe the measurement of the flux transmission field used in our correlation functions. We only discuss the most important steps; for a complete description of these processes and the DESI EDR Ly$\alpha$ catalogue, see \cite{cper}. 

In the observed wavelength regime we restrict our forests between $\lambda_\mathrm{min}=$3600 $\mathrm{\AA}$ and  $\lambda_\mathrm{max}=$5772$\mathrm{\AA}$. The lower limit is motivated by the limit of the DESI spectrograph, which for our Ly$\alpha$ catalogue translates to only including quasars above z$\sim$2. We set the upper limit to maximise the signal-to-noise of our correlation functions. In the rest-frame, our wavelength range is also reduced from the full extent of the Ly$\alpha$ forest to between $\lambda_\mathrm{RF, min}=$1040$\mathrm{\AA}$ and $\lambda_\mathrm{RF, max}=$1205$\mathrm{\AA}$, to avoid the wings of the Ly$\alpha$ and Ly$\beta$ emission lines, which can vary from quasar to quasar and thus add error to the continuum fitting. Compared to {\dMdB}, we have a higher upper wavelength limit in both the observed and rest-frame cases. In \cite{cper}, they found that $\lambda_\mathrm{RF, max}=$1205$\mathrm{\AA}$ was the limit corresponding to the lowest possible variance in the Ly$\alpha$ correlation function monopole, achieving a balance between the number of Ly$\alpha$ forest pixels and avoiding the wings of the Ly$\alpha$ emission line. 

The limits above result in 109\,900 Ly$\alpha$ forests, {with a maximum Ly$\alpha$ absorption pixel redshift of $z=3.75$}. {In practice we also place an upper limit on the quasar redshift of $z=3.75$ to be consistent with our mock catalogs \cite{hiram}, but this is done when computing the correlation functions and removes only a negligible number of forests.} For the Ly$\alpha$-quasar cross-correlation, we use a tracer catalog of quasars with a minimum redshift of $z>$1.77, containing 147\,899 quasars. This limit corresponds to a maximum possible separation of 300$h^{-1}$Mpc in the cross-correlation. In practice, we only measure correlations up to 200$h^{-1}$Mpc but scales up to 300$h^{-1}$Mpc are needed to accurately estimate the distortion (section \ref{sec:distortion_model}). We place final restrictions on our sample by rejecting forests with fewer than 50 valid \lyaf pixels, and forests with a failed continuum fit \cite{cper}. The resulting dataset contains 88\,509 Ly$\alpha$ forests.

Fluctuations around the mean transmitted flux fraction ($\delta_{\rm q}(\lambda)$) are computed as a function of wavelength as: 
\begin{equation}\label{eq:delta}
\delta_{\rm q}(\lambda) = \frac{f_{\rm q}(\lambda)}{\overline{F}(z) C_{\rm q}(\lambda)} - 1 ~,
\end{equation}
where $f_{\rm q}(\lambda)$ is the observed flux of quasar q, $\overline{F}(z)$ is the mean transmission of the IGM at redshift z ($z(\lambda) = \lambda/\lambda_{Ly\alpha} - 1$) and $C_{\rm q}(\lambda)$ is the unabsorbed quasar continuum. The product of the two terms in the denominator is equivalent to the mean expected flux for quasar q at observed wavelength $\lambda$. 

To measure the flux transmission field of each forest, we first estimate the product of the denominator terms. The first step is to make the approximation:
\begin{equation}\label{eq:delta_approx}
\overline{F}(z) C_{\rm q}(\lambda) = \overline{C}(\lambda_{\mathrm{RF}})\left( a_{\rm q} + b_{\rm q}\frac{\Lambda-\Lambda_{\mathrm{min}}}{\Lambda_{\mathrm{max}}-\Lambda_{\mathrm{min}}}\right) \hspace{0.5cm} \Lambda = \log\lambda ~,
\end{equation}
where $\overline{C}(\lambda_{\rm RF})$ is an estimate of the mean rest-frame spectra of all quasars, with the two quasar parameters ($a_{\rm q}$,$b_{\rm q}$) adjusting for quasar spectral diversity and the redshift evolution of the mean flux $\overline{F}(z)$. 

Using a first estimate of the continuum of each quasar, we characterise the flux variance ($\sigma_q^2$) in the pixels, later used to calculate the pixel weights (equation \ref{eq:lya_weight}), as the sum of two terms:
\begin{equation}\label{eq:sigma_forest}
\sigma^2_{\rm q} = \left[\overline{F} C_{\rm q}(\lambda)\right]^2 \sigma^2_\mathrm{LSS}(\lambda) + \eta(\lambda) \, \sigma^2_{\rm pip, q} ~.
\end{equation}
Both $\eta(\lambda)$ and $\sigma^2_\mathrm{LSS}(\lambda)$ functions are fitted iteratively with the mean quasar continuum $\overline{C}(\lambda_{\rm RF})$ and the per-quasar parameters  ($a_{\rm q}$,$b_{\rm q}$). 
We use the publicly available code \texttt{Picca}\footnote{https://github.com/igmhub/picca/} to compute these fits, and they typically converge in five iterations. The first term captures the intrinsic variance of the \lyaf, and the second term the instrumental noise variance. 
The noise variance estimated by the pipeline ($\sigma^2_{\rm pip,q}$) is normalized with the expected flux and is further multiplied by a function $\eta(\lambda)$ to correct for possible mis-calibrations of the instrumental noise.
In the {\dMdB} analysis, $\eta(\lambda)$ ranged from 1.05 and 1.2 between 3600 and 5500 $\mathrm{\AA}$ respectively. 
Over the same range in observed wavelength of the EDR+M2 dataset, this correction ranges from 1 to 1.01 \cite{cper}, demonstrating that the DESI pipeline gives a better noise estimate.

%% file: correlations.tex
\section{Measuring the correlation functions} \label{sec:correlations}
In this section we present the measurement of the correlation functions. We first describe the weights entering in the estimator of the correlation function in section \ref{sec:weights}, and discuss a small correction applied to the forest to model the distortion introduced by the continuum fitting (section \ref{sec:condis}). We then proceed to measure the auto-correlation of the \lyaf (section \ref{sec:autocorr}) and its cross-correlation with quasars (section \ref{sec:crosscor}), as well as the uncertainty in these measurements (section \ref{sec:covar}).

\subsection{Ly$\alpha$ weights} \label{sec:weights}

{Because of varying quasar brightness and exposure time, the signal-to-noise in our \lyaf is quite diverse. 
Therefore we weight the \lya pixels when measuring correlations. These weights are used to compute the auto-correlation in equation \ref{eq:autocorrcalc} and the cross-correlation in equation \ref{eq:wi_q}.
An optimal analysis would weight all \textbf{pairs} of pixels according to their inverse covariance matrix \cite{slosar2013,FontRibera_2018}.
However, this makes the analysis significantly more complex, so we follow previous analyses of eBOSS data ({\dMdB}) and weight pixels individually.}

Instead of directly using the inverse of the pixel variance (equation \ref{eq:sigma_forest}), we use the modified \lya weights introduced in \cite{cper}: 
\begin{equation} \label{eq:lya_weight}
 w_{\rm i}= \frac{\left[ (1+z_i)/(1+z_{\rm fid}) \right]^{\gamma_{\mathrm{Ly}\alpha}-1}}
 {\sigma^2_\mathrm{LSS}(\lambda) \sigma_\mathrm{mod}^2 + \eta(\lambda) \sigma^2_{\rm pip, q} / \left[\overline{F} C_q(\lambda)\right]^2} ~,
\end{equation}
{where $z_{\rm fid}$=2.25 (although this factor cancels in equation \ref{eq:autocorrcalc})}.
{The extra parameter $\sigma^2_{\rm mod}$ was introduced in \cite{cper} to modulate the relative importance of instrumental noise and intrinsic \lya fluctuations in the weights.
The value of $\sigma^2_{\rm mod}=7.5$ was found to be optimal for this dataset \cite{cper}, and it reduces the uncertainties on the auto- and cross-correlation by 20\% and 10\% respectively.}

{Finally, the numerator of equation \ref{eq:lya_weight} takes into account the redshift evolution of the \lya bias ($\gamma_{\mathrm{Ly}\alpha}$) and of the growth factor, and up-weights higher redshift pixels to favour absorption with higher signal. Following {\dMdB}, we use a value of $\gamma_{\mathrm{Ly}\alpha}=2.9$, motivated by the observation that the amplitude of the \lya power spectrum with redshift evolves as $(1+z)^{3.8}$ \cite{McDonald2006} .
This correction ranges from $\simeq$ 0.85 to 2.3 between $z=2$ and 4.}

\subsection{Continuum distortion}\label{sec:condis}

{As discussed in section \ref{sec:lyacat}, while fitting the quasar continua we fit two parameters per quasar ($a_{\rm q}$, $b_{\rm q}$) that modulate the amplitude and slope of the continuum in each forest.
However, this biases the mean and slope of each $\delta_{\rm q}(\lambda)$ toward 0 within each line of sight and distorts the measured correlations, often referred to as the continuum distortion \cite{Slosar_2011,B17}.
Following \cite{B17}, and to simplify the modelling of this distortion in section \ref{sec:distortion_model}, we modify the fluctuations $\delta_{\rm q}$ and make this distortion explicit:
\begin{equation}\label{eq:delta_eta}
    \hat{\delta}_{\rm i} = \sum\limits_{\rm i}\eta_{\rm ij}\delta_{\rm j} ~, 
\end{equation}
where the projection matrix $\eta_{ij}$ is given by
\begin{equation}\label{eq:proj_mat}
    \eta_{\rm ij}^{\rm q} = \delta^K_{\rm ij} - \frac{w_{\rm j}}{\sum\limits_{\rm k}{w_{\rm k}}} - \frac{w_{\rm j}\kappa_{\rm i}\kappa_{\rm j}}{\sum\limits_{\rm k} w_{\rm k}\kappa_{\rm k}^2} \hspace{0.5cm} \kappa_{\rm k} = \log\lambda_{\rm k} - \overline{\log\lambda_{\rm q}} ~,
\end{equation}
where $\delta^K_{\rm ij}$ is the Kronecker delta and weights $w$ are given in equation \ref{eq:lya_weight}.}

Finally, the transformation in equation \ref{eq:delta_eta} means that the average $\delta_{\rm q}$ at wavelength $\lambda$ is biased towards 0.
Following {\dMdB}, we subtract $\overline{\delta(\lambda)}$ when calculating the cross-correlation (section \ref{sec:crosscor}) such that:
\begin{equation}
    \tilde{\delta}_{\rm q}(\lambda) = \hat{\delta}_{\rm q}(\lambda) - \overline{{\delta}(\lambda)} ~.
\end{equation}
This correction guarantees that the cross-correlation (section \ref{sec:crosscor}) tends to 0 at large separations regardless of the quasar redshift distribution.

\subsection{Auto-correlation}\label{sec:autocorr}

To measure the correlation functions, redshift and angular ($\Delta z$,$\Delta \theta$) separations are first transformed to co-moving separations along ($r_\parallel$) and perpendicular ($r_\perp$) to the line of sight. 

We calculate ($r_\parallel,r_\perp$) as:
\begin{equation}
r_\parallel = [D_{\rm C}(z_{\rm i}) - D_{\rm C}(z_{\rm j})]\mathrm{cos}\left(\frac{\Delta\theta}{2}\right) ~, 
\end{equation}
and
\begin{equation}
r_\perp = [D_{\rm M}(z_{\rm i}) + D_{\rm M}(z_{\rm j})]\mathrm{sin}\left(\frac{\Delta\theta}{2}\right) ~,
\end{equation}
where (i,j) represent pixel-pixel or pixel-quasar pairs and $\Delta\theta$ the angle between them. 
The redshift of a pixel i is calculated assuming Ly$\alpha$ absorption, such that $z_{\rm i} = \lambda_{\rm OBS}/\lambda_{\rm {Ly}\alpha}-1$, where $\lambda_{\mathrm{Ly}\alpha}=1215.67\mathrm{\AA}$. The co-moving distance is given by $D_{\rm C}=c/100\int_0^z$ $\frac{dz'}{E(z')}$ in units of $h^{-1}$Mpc, where $E(z)=\sqrt{\Omega_{\mathrm{M}}(1+z)^3 + \Omega_{\Lambda}}$. $D_M$ is the transverse co-moving distance {(also known as angular co-moving distance)}, and $D_{\rm M}=D_{\rm C}$ \cite{Hogg}, as we assume $\Omega_{\rm k}$=0.

For the auto-correlation measurement we use a weighted covariance estimator following previous analyses \cite{Slosar_2011,dmdb2020}: 
\begin{equation}\label{eq:autocorrcalc}
\xi_{\rm A} = \frac{\sum_{\rm i,j \in A}w_{\rm i} w_{\rm j} \delta_{\rm i} \delta_{\rm j}}{\sum_{\rm i,j \in A} w_{\rm i} w_{\rm j}} ~,
\end{equation}
where A is a bin in ($r_\parallel,r_\perp$) space with width $4 h^{-1}$Mpc and the weights $w_{\rm i}$ are described in section \ref{sec:weights}. We sum over all pixels (i,j) across different lines of sight, but do not include pixels in the same line of sight because of continuum fitting errors that affect the entire forest, leading to spurious correlations which bias our measurement. In each bin A, our model correlation $\xi^{\rm mod}$ is evaluated at the weighted (by the number of pixel pairs) mean separation ($r_\parallel,r_\perp$) of the Ly$\alpha$ pixels (equation \ref{eq:lya_weight}) in our data. 

\begin{figure}
\hspace*{-2cm} 
    \centering  
    \includegraphics[scale=0.3]{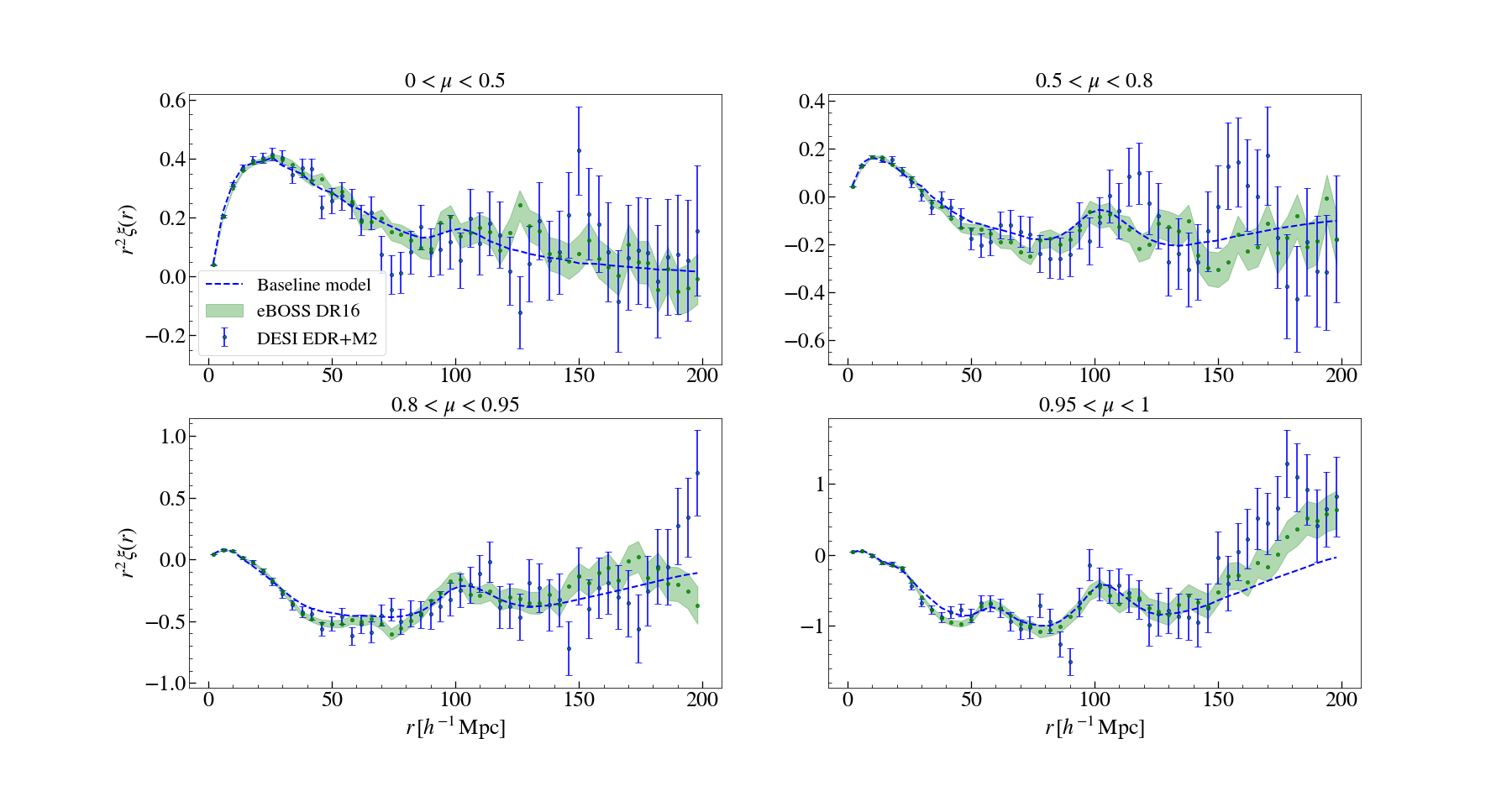}
    \caption{The DESI EDR+M2 (blue points) and eBOSS DR16 \cite{dmdb2020} (shaded green) Ly$\alpha$ auto-correlation compressed into weighted averages of $\mu$=$r_\parallel/r$, where $r=\sqrt{r^2_\parallel+r^2_\perp}$. We also include the best-fitting model to EDR+M2 described in section \ref{sec:modelling} (blue dashed). We have multiplied the correlation by $r^2$ to visualise the BAO peak, which is visible in both data sets. From these plots, we can see the consistency between our measurements and those in eBOSS DR16 - a validation of the quality of DESI data at an early survey stage. Note that the presence of three other bumps in the line of sight plot (bottom right) at 20, 60 and 111$h^{-1}$Mpc is due to correlations between the Ly$\alpha$ and SiIII(1207), SiII(1190)/SiII(1193) and SiII(1260) lines respectively. We model these contributions to the correlation in section \ref{sec:metals}.}
    \label{fig:wedge_auto}
\end{figure}

We measure the auto-correlation from $[0,200]h^{-1}$Mpc using 50 bins of 4$h^{-1}$Mpc along and perpendicular to the line of sight, giving us 2500 bins in total. The given limits on separation mean that the Ly$\alpha$ auto-correlation has $3.75\times10^{11}$ pixel pairs. In figure \ref{fig:wedge_auto} we present our Ly$\alpha$ auto-correlation measurement as a function of ($\mu,r$) where $\mu=r_{\parallel}/r$ and $r^2=r_\parallel^2+r_\perp^2$, for EDR+M2 and eBOSS DR16. We show the auto-correlation in 4 wedges, computed by averaging the 2D correlation in different selections of $\mu$, ranging from closest ($\mu\in[0.95,1]$) to furthest ($\mu\in[0,0.5]$) from the line of sight. In general, our DESI EDR+M2 measurement is consistent with {\dMdB}, and across all bins ($r_\parallel,r_\perp$), the uncertainties are only $\sim$1.9 times larger. {It should be noted that there is overlap between the two datasets, since DESI is re-observing quasars observed in {\dMdB}}. We present a constraint on the BAO peak amplitude from our best-fit model in section \ref{sec:fitter}. For $\mu \in [0.95.1]$, we can also see peaks caused by other atomic transitions from elements like Si, referred to as metal absorption or contamination in the rest of the paper. For example, we see the SiIII(1207) peak at $r\sim$20$h^{-1}$Mpc and the  SiII(1190)/SiII(1193) peak at $r\sim$60$h^{-1}$Mpc. At $r\sim100h^{-1}$Mpc we observe the BAO peak, which appears more prominently in the $\mu \in [0.95,1]$ plot, due to its overlap with the SiII(1260)x\lya peak at 111$h^{-1}$Mpc. In section \ref{sec:metals}, we discuss the origin of the metal peaks and how we model them.

\subsection{Cross-correlation}\label{sec:crosscor}

Similarly to the auto-correlation, we define an estimator for the cross-correlation as \cite{FontRibera2012,dmdb2020}:

\begin{equation}\label{eq:wi_q}
\xi_{\rm A} = \frac{\sum_{\rm i,j \in A}w_{\rm i} w_{\rm j} \delta_{\rm i}}{\sum_{\rm i,j \in A} w_{\rm i} w_{\rm j}} ~,
\end{equation}

\noindent for a pixel of Ly$\alpha$ absorption i and quasar j. In this case, the weights $w_{\rm j}$ are corrected for the quasar bias evolution and are given by: 

\begin{equation}\label{eq:qso_weight}
w_\mathrm{j} = \left(\frac{1+z_{\rm j}}{1+z_{\rm fid}}\right)^{\gamma_{\rm q}-1} ~,
\end{equation}

\noindent where $\gamma_{\mathrm{q}}$ = 1.44$\pm$0.08 \cite{dmdb2019}. {The $(1+z_{\rm fid})$ term, with $z_{\rm fid}$=2.25, cancels again in equation \ref{eq:wi_q}}. 
The sum in equation \ref{eq:wi_q} applies over all quasar-pixel pairs except pixels from their background quasar. Again, the correlation is binned in terms of line of sight and transverse separation ($r_\parallel,r_\perp$), with $r_\perp \in [0,200]h^{-1}$Mpc but now $r_\parallel \in [-200,200]h^{-1}\rm Mpc$. 
The cross-correlation is not symmetric under permutation of the two tracers, primarily because peaks from metal contamination appear at either $r_\parallel>$0 or $r_\parallel<$0, and because of systematic redshift errors (equation \ref{eq:drp}) and because the bias of each tracer evolves differently with redshift. Therefore we choose to define -ve separations for the case when the quasar is behind the Ly$\alpha$ pixel ($z_{\mathrm{q}}>z_{\mathrm{Ly}\alpha}$) and likewise +ve separations when the quasar is between the observer and the Ly$\alpha$ pixel ($z_{Ly\alpha}>z_{\rm q}$). This asymmetry allows us to study systematics like quasar redshift errors \cite{abault}. The given range of separation in both directions and the 4$h^{-1}$Mpc bin width now result in N = 100$\times$50 = 5000 bins.

\begin{figure}
\hspace*{-2cm} 
    \centering
    \includegraphics[scale=0.3]{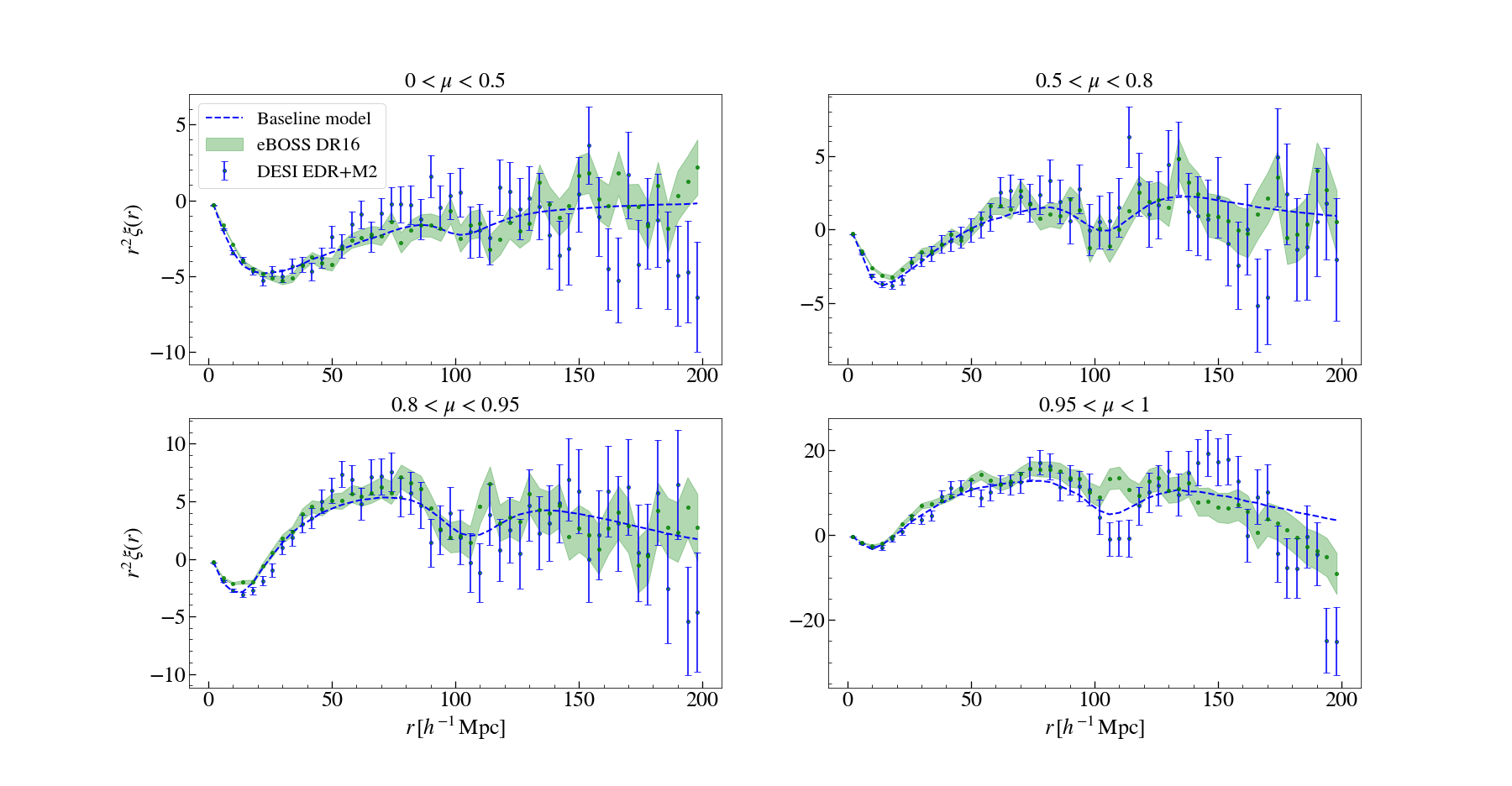}
    \caption{The 3D EDR+M2 (blue points) and eBOSS DR16 \cite{dmdb2020} (green shaded regions) Ly$\alpha$-quasar cross-correlation, and the baseline fit (blue dashed) to EDR+M2 described in section \ref{sec:modelling}. Because we have negative values of $r_\parallel$ (when $z_{\rm q}>z_{Ly\alpha}$), we have negative values of $\mu=r_\parallel/r$ and therefore average over $\mu\in[-1,1]$. The cross-correlation is expectedly noisier than the auto-correlation, but still we see a good level of consistency between eBOSS and DESI at this early stage. }
    \label{fig:wedge_cross}
\end{figure}

In figure \ref{fig:wedge_cross} we again show the cross-correlation as a function of $r$ and $\mu$, this time averaging over $\mu\in[-1,1]$, since the cross-correlation has -ve values of $r_\parallel$. Here, the correlation is reversed with respect to the matter correlation function. This behaviour arises from the fact that the bias of the Ly$\alpha$ forest is negative. By convention we trace the flux-transmission field, which is higher in more under-dense regions - and the quasar bias is positive, giving a negative product. The uncertainties in our DESI EDR+M2 cross-correlation measurement across all ($r_\parallel,r_\perp$) bins are $\sim$1.7 times larger than in eBOSS DR16. In the following sections, we discuss the covariance of the measurements of both the auto- and cross-correlations.

\subsection{Covariance matrix} \label{sec:covar}
We estimate the covariances of the auto- and cross-correlations using a sub-sampling method, following \cite{Delubac2015}. In this method, we define our sub-samples in 440 \texttt{HEALPix} \cite{Gorski_2005} pixels {using \texttt{Healpy}\cite{healpy}}, equivalent to an \textit{nside}=16, each of resolution 3.7 deg or solid angle 3.7$\times$3.7 = 13.4 deg$^2$. We then calculate the weighted covariance C$_\mathrm{AB}$ between samples as:
\begin{equation}\label{eq:covmat}
    {C_{\rm AB}} = \frac{1}{W_{\rm A}W_{\rm B}}\sum_{\rm i,j \in A}\sum_{{\rm k,l} \in B}{W^{\rm s}_{\rm A}W^{\rm s}_{\rm B}}\left[\hat{\xi}^{\rm s}_{\rm A}\hat{\xi}^{\rm s}_{\rm B} - \hat{\xi}_{\rm A}\hat{\xi}_{\rm B}\right] ~,
\end{equation}
where $\hat{\xi}^{\rm s}_{\rm A}$ is the measured correlation and $W^{\rm s}_{\rm A}$ is the summed weight in sub-sample s, and $W_{\rm A}=\sum_{\rm s}W^{\rm s}_{\rm A}$. We assign each Ly$\alpha$ pixel pair, or quasar-Ly$\alpha$ pair, to a given \texttt{HEALPix} pixel based on the object with the highest right ascension.
{Our effective redshift, $z_{\rm eff}=$2.376, is calculated by taking the average redshift of each bin of the correlation function weighted by the number of pixel pairs. The area of one \texttt{HEALPix} pixel at this redshift is $\sim$255$\times$255 ($h^{-1}$Mpc)$^2$}. The sub-sampling method of estimating covariance has also been validated in studies with mock catalogues in {\dMdB}.

Our auto- and cross-covariance matrices have N = 2500$\times$2500 = 6.25 million and N = 5000$\times$5000 = 25 million entries, with the diagonal entries dominating.

The normalised covariance, or correlation matrix, is defined as:

\begin{equation}\label{eq:corrmat}
    {Corr_{\rm AB}} = \frac{{C_{\rm AB}}}{\sqrt{Var_{\rm A}Var_{\rm B}}} ~,
\end{equation}

\noindent where ${Var_{\rm A}}$=${C_{\rm AA}}$ is the variance. The off-diagonal elements of equation \ref{eq:covmat} are noisy, and {we need to smooth them} to make our covariance matrices invertible. Following {\dMdB}, we model the off-diagonal elements of the correlation matrices as a function of line of sight and transverse separation, such that $Corr_{\rm AB}=Corr(\Delta{r}_\parallel$,$\Delta{r}_\perp)=Corr(|r_\parallel^{\rm A}-r_\parallel^{\rm B}|$,$|r_\perp^{\rm A}-r_\perp^{\rm B}|$). In figure 7 of {\dMdB} we see that the correlation matrix in this model decreases rapidly with ($\Delta{r}_\parallel$,$\Delta{r}_\perp$).

%% file: model.tex
\section{{Correlation function models}}\label{sec:modelling}
 
In this section, we discuss the theoretical model for our correlation functions, which includes the effect of all the major contaminants. For the most part, our model follows that in {\dMdB}, except the instrumental systematic effects model, which we adapt to account for the differences between the DESI and eBOSS instruments. {The final fit, which we describe in section \ref{sec:fitter}, consists of 13 free parameters shown in table \ref{table:bestfit}}.

\subsection{Power spectra}\label{sec:powerspec}

For our correlation model, we start with an isotropic linear power-spectrum template, decomposed into a peak and a smooth component. We then add the \lya/quasar Kaiser terms, some non-linear corrections, and contaminants, and convert the resulting anisotropic power spectrum into a correlation function. Finally, we multiply by the distortion matrix and recombine the peak and smooth components to obtain the final correlation model.

The auto- and cross-correlation functions are derived from the tracer biased power-spectrum, given by:
\begin{equation}\label{eq:plya}
P_{\mathrm{ij}}(k,\mu_k,z) = b_{\rm i}(z)b_{\rm j}(z)(1+\beta_i\mu_{\rm k}^2)(1+\beta_{\rm j}\mu_{\rm k}^2)P_{\mathrm{QL}}(k,\mu_{\rm k},z){F_\mathrm{NL,ij}(k,\mu_{\rm k})}G(k,\mu_{\rm k}) ~,
\end{equation}
where the vector ($k,\mu_k$) is defined such that $k^2 = k_\parallel^2+k_\perp^2$ and $\mu_k=k_\parallel/k$, {G is a term that accounts for binning on an ($r_\parallel,r_\perp$) grid (equation \ref{eq:G})} and $P_{\mathrm{QL}}$ is a quasi linear power spectrum {that is multiplied by a non-linear correction term $F_\mathrm{NL,ij}(k,\mu_k)$ (section \ref{sec:corrections})}. The parameters $b_i$ and $\beta_i$ are the linear bias and RSD of tracer i. 

To fit a unique function across each ($r_\parallel,r_\perp$) bin where the effective redshift varies from that of the full dataset ($z_{\rm eff}=2.376$), we need to evolve the Ly$\alpha$ and quasar biases using:
\begin{equation}\label{eq:biasshift}
    b_{\mathrm{Ly}\alpha}(z) = b_{\mathrm{Ly}\alpha} (z_{\rm eff}) \left(\frac{1+z}{1+z_{\rm eff}}\right)^{\gamma_{\mathrm{Ly}\alpha}} ~,
\end{equation}
\begin{equation}
    b_{\mathrm{q}}(z) = b_{\mathrm{q}} (z_{\rm eff}) \left(\frac{1+z}{1+z_{\rm eff}}\right)^{\gamma_{\mathrm{q}}} ~,
\end{equation}
where $\gamma_{\mathrm{Ly}\alpha}=$2.9 \cite{McDonald2006} and $\gamma_{\mathrm{q}}=$1.44\cite{dmdb2019}.

For the auto-correlation (i=j) the only tracer is Ly$\alpha$ absorption and for the cross (i$\neq$j) the tracers are Ly$\alpha$ absorption and quasars. Following past BAO analyses ({\dMdB}), we take the linear matter power spectrum from Planck 2018\footnote{derived using \texttt{CAMB} https://github.com/cmbant/CAMB} \cite{Planck2018} and separate it into a smooth component and peak component with non-linear broadening corrections: 
\begin{equation}\label{eq:pksmooth}
P_{\mathrm{QL}}(\vec{k},z) = P_{\mathrm{peak}}(k,z)\mathrm{exp}\left[-\frac{k_\parallel^2\Sigma_\parallel^2 + k_\perp^2\Sigma_\perp^2}{2}\right] + P_{\mathrm{smooth}}(k,z) ~,
\end{equation}
where $P_{\mathrm{peak}}$ contains BAO and $P_{\mathrm{smooth}}$ does not. We make this decomposition using the side band technique from \cite{Kirkby2013} such that $P_{\mathrm{peak}} = P_{\mathrm{lin}}-P_{\mathrm{smooth}}$, and allows us to isolate the model for the BAO peak from the rest of the correlation and fit its position and amplitude. For the non-linear broadening correction of the BAO peak, we use {$\Sigma_\parallel=6.36 h^{-1}$Mpc and $\Sigma_\perp=3.24h^{-1}$Mpc} as in {\dMdB} \cite{dmdb2020,Eisenstein2007}, assuming that our small difference in effective redshift has negligible effect.

The last term in equation \ref{eq:plya}, $G$, corrects for the fact that we bin correlations in a grid in ($r_\parallel,r_\perp$) and is the product of the Fourier transform of two top-hat functions:
\begin{equation}\label{eq:G}
G(k_\parallel,k_\perp) = \mathrm{sinc}\left(\frac{k_\parallel \Delta{{A}_\parallel}}{2}\right)\mathrm{sinc}\left(\frac{k_\perp \Delta{{A}_\perp}}{2}\right) ~,
\end{equation}
where $\Delta{{A}_\parallel}=4h^{-1}$Mpc and $\Delta{{A}_\perp}=4h^{-1}$Mpc are the bin widths in the line-of-sight and transverse directions respectively. To compute equation \ref{eq:G}, we make the approximation that our correlation is distributed homogeneously within each bin. In reality, it scales with $r_\perp$ in the transverse direction, but \cite{B17} have shown that the approximation made here produces an accurate correlation function.

\subsection{Non-linear corrections}\label{sec:corrections}

We multiply the quasi-linear power spectrum $P_{\mathrm{QL},\mathrm{ij}}(k,\mu_k)$ of equation \ref{eq:plya} by a non-linear correction $F_\mathrm{NL,ij}(k,\mu_k)$. In the auto-correlation, the dominant effects encompassed by this correction are non-linear structure growth, peculiar motion, and broadening due to the thermal motion of particles, modelled in \cite{Arinyo_i_Prats_2015}.

{Statistical redshift errors and galaxy peculiar velocities of quasars smear the cross-correlation along the line of sight $r_\parallel$}, which following \cite{Percival_2009} we model as Lorentzian damping:

\begin{equation}\label{eq:sigma_v_qso}
F_{\rm NL}^{\rm cross}(k_\parallel) = \frac{1}{\sqrt{1+(k_\parallel \sigma_{\rm v})^2}} ~,
\end{equation}

\noindent where $\sigma_{\rm v}$ is one of the free parameters in our model, given in table \ref{table:bestfit}. {Statistical errors also affect the auto-correlation by adding random shifts to quasar spectral templates that in turn introduce systematic error in the mean continuum estimate $\overline{F}(z)C_q(\lambda)$ (equation \ref{eq:delta_approx}). The impact of statistical redshift errors on BAO measurements from both the auto- and cross-correlation is described in \cite{syoules}.}

\subsection{Modeling contaminants}

Both the auto- and cross-correlation functions receive small contributions from other effects, which for the auto-correlation we model as: 

\begin{equation}\label{eq:auto_f}
\xi_{\mathrm{auto}} = \xi_{\mathrm{Ly}\alpha \times \mathrm{Ly}\alpha} + \sum_\mathrm{m} \xi_{\mathrm{Ly}\alpha \times \mathrm{m}} + \sum_{\mathrm{m_i,m_j}} \xi_{\mathrm{m_i} \times \mathrm{m_j}} + \xi_{\mathrm{inst}} ~,
\end{equation}

\noindent where $\xi_{Ly\alpha \times Ly\alpha}$ is transformed from $P_{\rm Ly\alpha}(k,\mu_k)$ (equation \ref{eq:plya}) via a Fast Hankel transform. The \lya bias parameter is also modelled to account for contamination by high column density (HCD, see section \ref{sec:biaspars}) systems. $\xi_{Ly\alpha \times m}$ is the cross-correlation between $Ly\alpha$ absorption and metal absorption and $\xi_{m_i \times m_j}$ is the metal-metal auto-correlation discussed in section \ref{sec:metals}. Finally, $\xi_{\mathrm{inst}}$ is the contribution from DESI instrumental systematics, including sky subtraction residuals and spectro-photometric calibrations, described in full in \cite{Satya} and modelled in section \ref{sec:specphoto}. Likewise, for the cross-correlation, we write:

\begin{equation}\label{eq:cross_f}
\xi_{\mathrm{cross}} = \xi_{\mathrm{Ly}\alpha \times \mathrm{q}} + \sum_\mathrm{m} \xi_{\mathrm{q} \times \mathrm{m}}  + \xi^{\mathrm{TP}} ~,
\end{equation}

where $\xi_{\mathrm{Ly}\alpha \times \mathrm{q}}$ is transformed from Fourier space in the same way as the auto-correlation, $\xi_{\mathrm{q} \times \mathrm{m}}$ is the quasar-metal line cross-correlation and $\xi^{\mathrm{TP}}$ is the contribution from the effect of quasar radiation given in section \ref{sec:qsorad}. 

Finally we introduce another free parameter to our model to account for any systematic shift in line-of-sight separation between quasars and Ly$\alpha$ absorption:

\begin{equation}\label{eq:drp}
    \Delta\mathrm{r}_{\parallel,\mathrm{q}} = r_{\parallel,\mathrm{True}} - r_{\parallel,\mathrm{Measured}} ~.
\end{equation}

\noindent This effect arises due to systematic quasar redshift errors and thus produces an asymmetry over positive and negative r$_\parallel$ which can be studied using the cross-correlation of quasars and Ly$\alpha$ absorption, as is done in \cite{abault}.

\subsubsection{Quasar radiation effects}\label{sec:qsorad}

At small scales, quasar radiation can strongly affect its surrounding gas or a nearby Ly$\alpha$ forest. This radiation dominates over the UV background \cite{dmdb2017}, so the ionisation fraction, and therefore the number of passing Ly$\alpha$ photons, increases. This is modelled following \cite{FontRibera2013,dmdb2020}, assuming isotropic quasar emission: 

\begin{equation}\label{eq:tp_eff}
    \xi^{\mathrm{TP}} = \xi^{\mathrm{TP}}_0 \left(\frac{1 h^{-1}\mathrm{Mpc}}{r}\right)^2 \exp\left(\frac{-r}{\lambda_{\mathrm{UV}}}\right) ~,
\end{equation}

\noindent where we fit for the scale parameter $\xi^{\mathrm{TP}}_0$, and $\lambda_{\mathrm{UV}}$ is the UV photon mean free path set to 300$h^{-1}$Mpc following \cite{Rudie2013}.

\subsubsection{Instrumental systematics}\label{sec:specphoto}
{The focal plane of DESI is divided in 10 "petals", with 500 fibers each that are fed to one of the 10 spectrographs. The calibration of DESI spectra (sky subtraction using sky fibers and flux calibration using standard stars) are performed separately for each of the 10 petals \cite{JGuy}. This introduces a possible source of systematic correlations for pairs of spectra obtained with the same spectrograph. The impact of correlated sky residuals and flux calibration errors in the \lya correlations in DESI is described in detail in \cite{Satya}. Poisson fluctuations at the observed wavelength of the sky lines within each spectrograph, and flux calibration residuals, produce an excess correlation at $r_\parallel=0$. We model the contributions of both of these, which contribute roughly equally, in the auto-correlation with an empirical functional form:}
\begin{equation}
    \xi_{\mathrm{inst}} = 
        \begin{cases}
          {\rm A}_{\mathrm{inst}}\left(\frac{r_\perp}{80} - 1\right)^2 & r_\perp<80 h^{-1}\mathrm{Mpc}, ~r_\parallel=0\\
          0 & r_\perp>80 h^{-1}\mathrm{Mpc}\\
        \end{cases} 
\end{equation}
where A$_\mathrm{inst}$ is fit and given in section \ref{sec:fitter}. {The limit of 80$h^{-1}$Mpc roughly corresponds to the angular size of a DESI petal on the sky at the effective redshift of our data}. {Note that we do not see this effect in the cross-correlation since it requires two correlating subtracted sky lines in two \lya forests.}
In {\dMdB} the sky subtraction residuals were modelled with a Gaussian with two free parameters instead of one. The differences in the DESI and eBOSS instruments - such as eBOSS having only two spectrographs with 500 fibres each - lead to a difference in the shape and amplitude of the instrumental effects model. In section \ref{sec:fitter}, we compare the amplitude of both models at $r_\parallel=r_\perp=0$. 

\begin{figure}
\hspace*{-1.5cm} 
    \centering
    \includegraphics[scale=0.4]{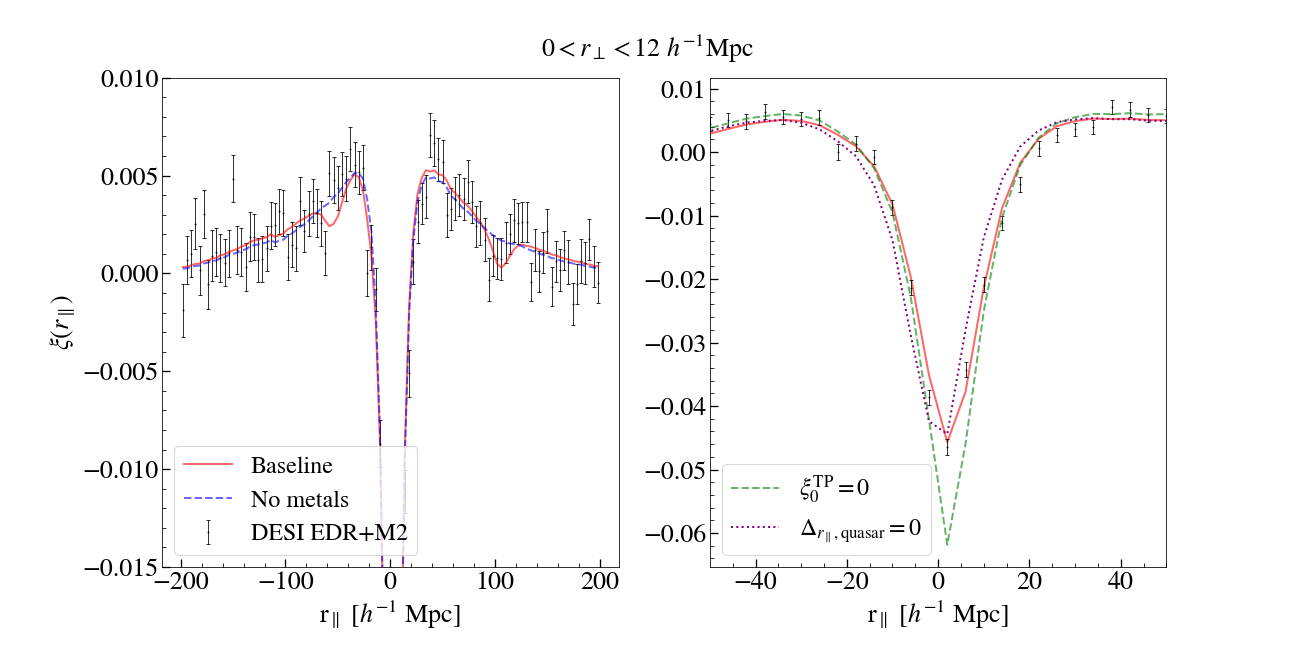}
    \caption{The \lya-quasar cross-correlation as a function of $r_\parallel$, for ${r_\perp}{\in}[0,12]h^{-1}$Mpc. On the right, we over-plot our baseline model (red solid) reported in table \ref{table:bestfit}  and a  model with no metal line component (blue dashed). On the left, we show two additional lines to highlight the effect of setting $\Delta{r}_{\parallel,\mathrm{q}}$ to 0 and ignoring quasar radiation effects (green dashed). We can see the asymmetry of the cross-correlation through metals and systematic redshift errors (parameterised by $\Delta{r}_{\parallel,\mathrm{q}}$) in the data here.
    }
    \label{fig:1d_cross_fit}
\end{figure}

\subsubsection{Absorption by high column density systems} \label{sec:biaspars}
The Ly$\alpha$ forest contains absorption from neutral hydrogen in the diffuse IGM and systems of neutral hydrogen with significantly higher column density{\footnotemark}. In section \ref{sec:pixmask}, we defined DLAs as having column density $N_{HI}>10^{\mathrm{20.3}}$cm$^{-\mathrm{2}}$ and discussed how we identify and account for these using pixel masks and a Voigt profile to model the wings. Assuming we identify and treat all of the DLAs, the noise in our correlation measurements should be significantly reduced. We still expect to be contaminated by some objects of column density $N_{HI}<$10$^{{20.3}}$cm$^{-{2}}$ (sometimes referred to as sub-DLAs) which have a similar absorption profile, but lesser effect on our data. Ideally, we would also mask HCDs below the DLA limit, but these are difficult to identify with our current pipeline and resolution. Following \cite{FontRibera2012,Rogers2018} we model the contamination from these systems as a modification of the large-scale bias parameters:

\footnotetext{\textit{High column density systems} is a term used in past eBOSS publications to refer to neutral hydrogen with column density $N_{HI}>10^{\mathrm{17.2}}$cm$^{-\mathrm{2}}$, of which DLAs are the highest column density subset and the most impactful contaminants on our data.}
\begin{equation}
    \tilde{b}_{\mathrm{Ly}\alpha} = b_{\mathrm{Ly}\alpha} + b_{\mathrm{HCD}}F_{\mathrm{HCD}}(k_\parallel) ~,
\end{equation}
\begin{equation}
    \tilde{b}_{\mathrm{Ly}\alpha}\tilde{\beta}_{\mathrm{Ly}\alpha} = b_{\mathrm{Ly}\alpha}\beta_{\mathrm{Ly}\alpha} + b_{HCD}\beta_{\mathrm{HCD}}F_{\mathrm{HCD}}(k_\parallel) ~,
\end{equation}
where $b_{\mathrm{HCD}}$ and $\beta_{\mathrm{HCD}}$ are the bias and RSD parameter of the HCD systems and $F_{\mathrm{HCD}}(k_\parallel)$ is a factor depending on the distribution of HCDs, modelled following {\dMdB} as:
\begin{equation}
    {F}_{\mathrm{HCD}}(k_\parallel) = \exp(-{L}_{\mathrm{HCD}}k_\parallel) ~.
\end{equation}
Here $L_{\mathrm{HCD}}$ is the typical HCD length scale, and is set to 10$h^{-1}$Mpc following {\dMdB} because of its degeneracies with other parameters in the fit. Note that it was verified in \cite{Cuceu2022_mocks,Cuceu_Bayes} that varying $L_{\mathrm{HCD}}$ between $\sim$7-13$h^{-1}$Mpc does not significantly affect BAO measurements.

\subsubsection{Absorption by metal lines}\label{sec:metals}
As stated earlier and shown in equations \ref{eq:auto_f} and \ref{eq:cross_f}, our measured correlations are not only the result of Ly$\alpha$ absorption in the IGM or HCDs but also heavier elements in the IGM due to galactic gas outflows \cite{Yang2022}. Therefore, we introduce a cross power spectrum P$_{\mathrm{mn}}(\mathbf{k},z)$ for each pair of absorbers (m,n) and their configuration-space counterparts $\xi^{\rm m \times n}$. 

Because we measure our flux-transmission field and correlations assuming all absorption in the Ly$\alpha$ is due to the Ly$\alpha$ transition, we assign to each absorption pixel a redshift $z=\lambda_{\mathrm{obs}}/\mathrm{1215.67{\AA}} - 1$. However, because of the metal lines from heavier elements, {an excess correlation} appears at the separation between the true and assigned redshifts $z_\mathrm{m}$ and $z_{\mathrm{Ly}\alpha}$:

\begin{equation}
    r_\parallel = d_{\mathrm{m}} - d_{\mathrm{Ly}\alpha} = c\int_{z_{\mathrm{Ly}\alpha}}^{z_{\mathrm{m}}} \frac{dz'}{H(\overline{z})} \approx \frac{c(1+z)}{H(z)}\frac{\lambda_{\rm m} - \lambda_{\mathrm{Ly}\alpha}}{\overline{\lambda}} ~,
\end{equation}

\noindent where $\overline{\lambda}$ and $\overline{z}$ are the mean values of the Ly$\alpha$ and metal absorption. 
{The excess correlation is more clearly observed for vanishing physical separations between two absorbers, when $r_\perp\sim0$}. We represent the correlation shift for each pair of absorbers (m,n) using the metal matrix formalism of \cite{Blomqvist2018}. In this method, the shifted correlation function $\xi^{\mathrm{m} \times \mathrm{n}}_1$ for each absorber pair (m,n) is given in terms of the un-shifted correlation function $\xi^{\mathrm{m} \times \mathrm{n}}_0$ as:

\begin{equation}
    \xi^{\mathrm{m} \times \mathrm{n}}_1 (\mathrm{A})= \sum_{\mathrm{B}} M_{\mathrm{AB}},\xi^{\mathrm{m} \times \mathrm{n}}_0(r_\parallel(\mathrm{B}),r_\perp(\mathrm{B})),
\end{equation}

\noindent where (m,n)$\in$B is the pixel separation using the rest wavelength of each absorber m and n, (m,n)$\in$A refers to the pixel separation assuming both pixels have rest wavelength $\lambda_{\mathrm{Ly}\alpha}$, and:

\begin{equation}
    M_{\mathrm{AB}} =\frac{\sum_{\mathrm{(m,n)} \in \mathrm{A}, \mathrm{(m,n)} \in \mathrm{B}} w_\mathrm{m}w_\mathrm{n}}{\sum_{\mathrm{(m,n)} \in \mathrm{A}} w_\mathrm{m}w_\mathrm{n}} ~.
\end{equation}

\noindent For the cross-correlation, index $m$ is a quasar with redshift $z_{\rm q}$ used to calculate the projected separation. To compute $M_{\rm AB}$ in a reasonable time, we only use 1\% of pixel pairs. We have verified that this does not affect the measured metal biases.

{Different rest-frame wavelengths peaks at different line of sight separations. Given the range of separations that we measure, we model 4 dominant metal lines - SiII(1190), SiII(1193), SiIII(1207), and SiII(1260) \cite{B17}.} Each metal line has it's own bias and RSD parameters ($b_\mathrm{m},\beta_\mathrm{m}$) and corresponding power spectrum P$_\mathrm{mn}$ as mentioned above. We cannot determine these two parameters separately since the correlations $\xi^{\mathrm{m} \times \mathrm{n}}_1$ only have a significant impact at small $r_\perp$, and therefore we set $\beta_\mathrm{m}=$0.5 following \cite{B17}. The values of $b_{\mathrm{m}}$ for each metal line are constrained in the fits discussed in section \ref{sec:fitter}. {The CIV metal bias has also been typically fit in previous Ly$\alpha$ forest correlation studies \cite{Blomqvist2018,SatyaCIV}, but we decide not to model it in this study because it was weakly detected and had a negligible effect on the {\dMdB} analysis, and we report the same for EDR+M2 data.}

\subsection{Modelling the continuum distortion} \label{sec:distortion_model}

Due to distortions created in the quasar continuum fitting process, we apply two corrections to our measured delta fields in section \ref{sec:lyacat}. Doing this, each measured delta $\tilde{\delta}_{\rm q}(\lambda)$ becomes a linear combination of all other pixels in the same forest, distorting the correlation functions measured from these deltas. To correct for the distortion we relate the measured ($\hat{\xi}$) and true ($\xi^{\rm t}$) correlations following \cite{B17,dmdb2017,dmdb2020} with:
\begin{equation}\label{eq:distortion_transform}
\hat{\xi}_{\rm A} = \sum_{\rm A'} D_{\rm AA'} \xi^{\rm t}_{\rm A'} ~,
\end{equation}
where 
\begin{equation}\label{eq:dmat}
    D_\mathrm{AA',auto} = \frac{\sum\limits_{({\rm i,j})\in {\rm A}}w_{\rm i}w_{\rm j}\left(\sum\limits_{\rm (i',j' \in A')} \eta_{\rm ii'}\eta_{\rm jj'}\right)}{\sum\limits_{\rm (i,j)\in A}w_{\rm i}w_{\rm j}} ~,
\end{equation}
for the auto-correlation, where $\eta$ is the projection matrix given in equation \ref{eq:proj_mat}. We also have
\begin{equation}\label{eq:xdmat}
    D_\mathrm{AA',cross} = \frac{\sum\limits_{\rm (i,k) \in A}w_{\rm i} \sum\limits_{\rm (j,k) \in A'}\eta_{\rm ij}}{\sum\limits_{\rm (i,k) \in A}w_{\rm i}} ~,
\end{equation}
for the cross-correlation, where $\rm{i,j}$ are pixels in the same forest and $\rm k$ is a quasar. Like the metal matrices, we compute the distortion matrices with only 1\% of total pixel pairs to reduce computing hours. In our analysis, we model the effect of distortion out to a separation of 300$h^{-1}$Mpc, since distortions at this scale are known to affect the correlation function at scales relevant to our analysis. The two expressions above equate to multiplying our physical model of the correlation by the distortion matrix before any comparison with data. The effects of distortion have also been validated in previous studies using mocks \cite{B17}.

%% file: fitting.tex
\section{Fits to the data}\label{sec:fitter}
We now present the best-fitting solution of the model introduced in section \ref{sec:modelling} to the \lya auto-correlation and \lya quasar cross-correlation described in section \ref{sec:correlations}. For this study, we have established a baseline model with 13 free parameters based on Kaiser models for biased tracers with additional contaminants, described in detail in section \ref{sec:modelling}. 

In the anisotropic correlation model of equation \ref{eq:plya}, it is evident that the cross-correlation we measure is only sensitive to the product of the Ly$\alpha$ and quasar biases. As a result, we present only the auto and auto+cross combined fits. In the latter, we break the bias degeneracy to measure the quasar bias b$_\mathrm{q}$ distinctly. 
Then, using the fact that for point tracers $\beta_\mathrm{q}$=$f/b_\mathrm{q}$, where $f=0.97$ \cite{Planck2018} is the growth-rate {for our fiducial cosmology at $z=2.376$}, we infer a value of $\beta_\mathrm{q}$. 
The other advantage of combining both correlations is improving constraints on key parameters. For example, our constraints on A$_\mathrm{BAO}$ improve by 1.6x going from auto-correlation fit alone to the combined fit.

In table \ref{table:bestfit}, mean values and 68\% credible regions of the posterior distributions for each free parameter in the auto and combined auto+cross correlations are given\footnote{computed using the plotting tool \texttt{Getdist}\ \cite{getdist}}. To perform each fit, we use the Python package \texttt{Vega}\footnote{https://github.com/andreicuceu/vega}, which employs the nested sampler \texttt{Polychord} \cite{polychord} to sample each parameter. We find that the auto-only results are very consistent with the auto+cross combined. In each column we put a "-" symbol in entries where parameters are only relevant to the combined correlation and not to the auto-correlation alone. 

The second part of table \ref{table:bestfit} gives the characteristics of each fit, including the number of bins in each measurement, the number of free parameters, the minimum $\chi^2$ value, and the probability (p-value). Note that to obtain the latter two values, we use a minimiser instead of the nested sampler used to obtain parameter means and confidence regions. {The minimiser makes use of the Python package \texttt{Iminuit}\footnote{https://github.com/scikit-hep/iminuit}} We perform the fit for $r\in[10,180]h^{-1}$Mpc due to the increasing complexity of modelling the correlation function at small scales (non-linearities). The 4$h^{-1}$Mpc bin width of the correlation functions translates to 1590 and 4770 bins for the auto and combined fits.

\begin{table}
\centering
\begin{tabular}{lllll}
\hline
\hline
                 Parameter & Auto & Combined \\
\hline
\hline
    $\mathit{b}_{\mathrm{Ly}\alpha}$   & -0.129$\pm$0.010 &  -0.134$\pm$0.009  \\
    $\beta_{\mathrm{Ly}\alpha}$ &   1.49$^{+0.14}_{-0.19}$  &   1.41$^{+0.12}_{-0.15}$   \\
    $\mathit{b}_{\mathrm{HCD}}$ & -0.045$\pm$0.009  &   -0.039$\pm$0.009    \\
	$10^3 \mathit{b}_{\eta,\mathrm{SiII(1190)}}$ &  -3.3$\pm$1.0  &     -2.2$\pm$0.8 \\
	$10^3 \mathit{b}_{\eta,\mathrm{SiII(1193)}}$ & -1.7$^{+1.1}_{-0.8}$ & -0.9$^{+0.8}_{-0.3}$\\
	$10^3 \mathit{b}_{\eta,\mathrm{SiII(1260)}}$ &    -2.9$\pm$1.1   &    -2.6$\pm$0.9 \\
	$10^3 \mathit{b}_{\eta,\mathrm{SiIII(1207})}$ &     -3.9$\pm$1.0 & -3.4$\pm$0.9 \\
	$\mathit{b}_{\mathrm{q}}$    &   -         &     3.41$\pm$0.16   \\
	$\Delta{\mathit{r}}_{\parallel,\mathrm{q}} (h^{-1}\mathrm{Mpc})$ &   - &   -2.21$\pm$0.18 \\
	$\sigma_{\rm v} (h^{-1}\mathrm{Mpc})$ &        -    &   5.2$\pm$0.5 \\
	$\xi^{\mathrm{TP}}_0$ &    -    &               0.68$\pm$0.18\\
	$10^4 \mathit{A}_{\mathrm{inst}}$ &  2.6$^{+0.4}_{-0.8}$     &         2.4$^{+0.3}_{-0.5}$ \\
    A$_{\mathrm{BAO}}$ & 1.04$\pm$0.45  & 1.21$\pm$0.32 \\
\hline
\hline
 N$_{\mathrm{bin}}$ &    1590   &  4770     \\
 N$_{\mathrm{params}}$  &   9    &    13       \\
     $\chi^2_{\mathrm{min}}$  &  1711    &   4945          \\
    probability    &  0.01    &    0.03     \\
     
\hline
\hline
\end{tabular}
\caption{Table of early DESI fits on the auto-correlation and the combined auto+cross correlations. We do not perform a cross-correlation-only fit due to degeneracies between parameters. In the first part we show the mean and 68\% credible region for each free parameter in our model computed from the posterior distributions from the nested sampler. We evaluate each parameter at the effective redshift of our data $z_{\mathrm{eff}}=$2.376, where redshift is relevant. Where a parameter does not feature in the auto-correlation fit, we use the "-" symbol. In the second part, we give the characteristics of the fit, including the number of bins, free parameters, and the $\chi^2_{\mathrm{min}}$ probability as estimated with a minimiser.}
\label{table:bestfit}
\end{table}

To measure the full correlation function we re-combine the peak and smooth parts of the correlation (outlined in section \ref{sec:powerspec}):
\begin{equation}\label{eq:pk_smooth}
\xi(r_\parallel,r_\perp,\alpha_\parallel,\alpha_\perp) = \xi_{\mathrm{smooth}}(r_\parallel,r_\perp) + \mathrm{A_{BAO}}\xi_{\mathrm{peak}}(\alpha_\parallel r_\parallel,\alpha_\perp r_\perp) ~,
\end{equation}
where $\alpha_\parallel$ and $\alpha_\perp$ are the BAO parameters {(see {\dMdB} for recent constraints on these)}. Here $\mathrm{A_{BAO}}$ is the amplitude of BAO, which we detect at 3.8$\sigma$ confidence (table \ref{table:bestfit}). We fix both $\alpha_\parallel$ and $\alpha_\perp$ to 1, or the fiducial cosmology, {and focus on preparing for a robust measurement with DESI year 1 data.}

\begin{figure}
\hspace*{-0.5cm} 
    \centering  
    \includegraphics[scale=0.5]{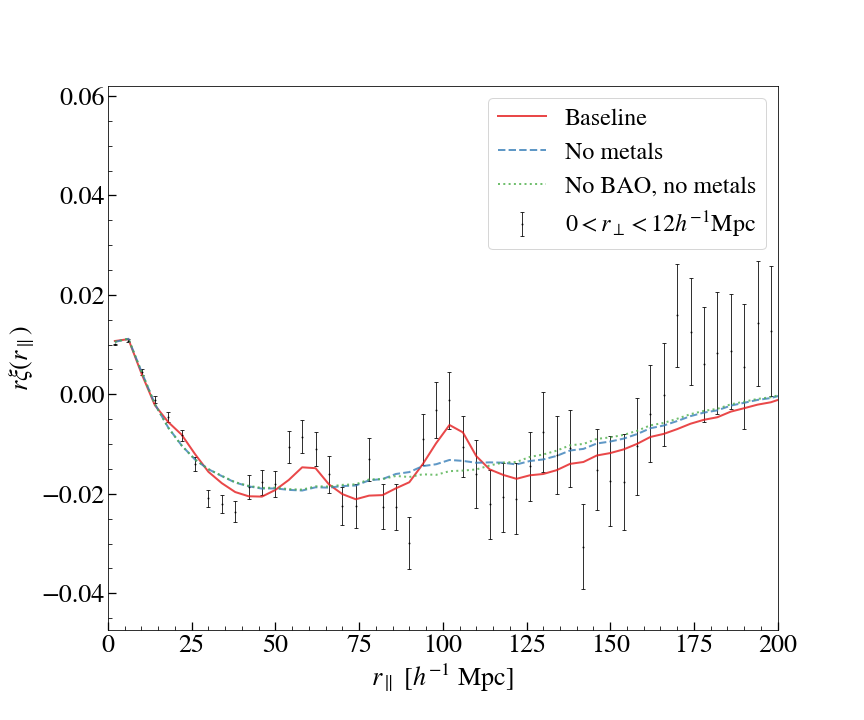}
    \caption{{The best fitting solution (red solid) of our model to the DESI EDR+M2 \lya auto-correlation (black points) function along the line of sight, where we average over 0$<r_\perp<12h^{-1}$Mpc. We also include a fit which does not model metal contamination (blue dashed) and a fit which does not model metal contamination or BAO (green dotted). We multiply the y-axis by r for visualisation. The bumps in correlation caused by SiIII(1207) and SiII(1190)/SiII(1193) at r$_{\parallel}\sim$20$h^{-1}$Mpc and r$_{\parallel}\sim$60$h^{-1}$Mpc are visible. The most prominent peak is driven by SiII(1260) at r$_{\parallel}\sim$111$h^{-1}$Mpc, rather than the overlapping BAO peak at $r_{\parallel}$=100$h^{-1}$Mpc.}}
    \label{fig:auto_wedge_high}
\end{figure}

The best-fitting model is shown on four auto-correlation wedges in figure \ref{fig:wedge_auto} and the same four cross-correlation wedges in figure \ref{fig:wedge_cross}. In figure \ref{fig:auto_wedge_high}, we also show the auto-correlation along the line of sight, averaged over 0$<r_\perp<12h^{-1}$Mpc and overlaid with the baseline model from the minimiser fit. We also include a fit that does not model the effect of metal contamination, and a fit that does not model metal contamination or BAO. We multiply by $r$ to see that the bumps caused by the metal lines SiIII(1207) and SiII(1190)/SiII(1193) at $r\sim$20h$^{-1}$Mpc and $r\sim$60$h^{-1}$Mpc respectively are well-fit by our model, and poorly-fit by the no metal model. The SiII(1260) line gives a produces a peak at $r\sim$111$h^{-1}$Mpc, which along the line of sight dominates over the BAO peak with which it overlaps.

\subsection{Fit probability}
The combined fit presented in table \ref{table:bestfit} has a $\chi^2$ probability of 0.03. Because this is a relatively low value, we discuss several caveats here.

\begin{figure}
\hspace*{-0cm} 
    \centering  
    \includegraphics[scale=0.5]{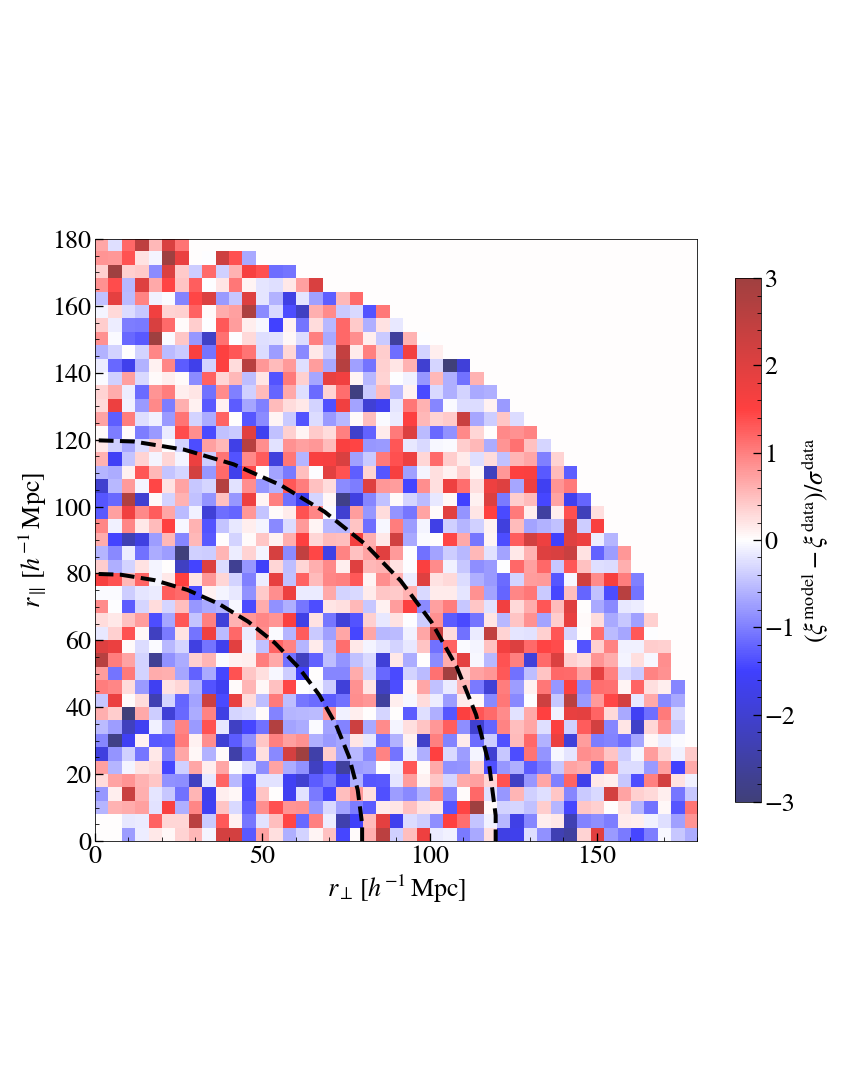}
    \vspace*{-30mm}
    \caption{Residuals of our fit to the DESI EDR+M2 auto-correlation, defined as the difference between data and model correlation divided by the uncertainties of each data point (square root of the diagonal of the covariance matrix). The area between the black dashed lines indicates the BAO region between 80-120$h^{-1}$Mpc. We see that overall there are no strong areas of discrepancy.}
    \label{fig:2dresidauto}
\end{figure}

In figure \ref{fig:2dresidauto}, the residuals are spread evenly across ($r_\parallel$,$r_\perp$). There are some groups of bins with higher residuals closer to the line-of-sight (r$_\perp=0h^{-1}$Mpc) {where continuum distortions are strongest,} and at larger separations $r\sim150h^{-1}$Mpc, but these are likely due to the high degree of correlation between data points which we ignore in this plot (we use the diagonal of the covariance matrix here).

{If we change the lower limit on our fit of both correlations to $r=20h^{-1}$Mpc rather than 10$h^{-1}$Mpc, our fit probability increases to 0.08 from 0.03. It's possible that our constraints on the baseline model are slightly biased when fitting down to 10$h^{-1}$Mpc, but we will leave this mock catalogues studies in DESI year 1. Recent results using \lya mocks on \texttt{AbacusSummit} \cite{BoryanaAbacus} also suggest that the Kaiser model that we employ here does not adequately model non-linear effects in the \lya quasar cross-correlation for $r<30h^{-1}$Mpc. If we set this as a lower limit on the cross-correlation while keeping 10$h^{-1}$Mpc for the auto-correlation (\cite{BoryanaAbacus} find the Kaiser model is a good approximation in this case), we increase our probability to 0.05, without biasing our measurements. 

Changing the upper limit to $r=150h^{-1}$Mpc also increases the fit probability slightly to 0.05, and has a negligible effect on measured parameter values. For all of these tests, the most important thing for future analyses will be the effect on the BAO scale parameters, which we do not measure here. Therefore, we keep the baseline analysis between r$\in[10,180]h^{-1}$Mpc as in {\dMdB}}.

\footnotetext{The Ly$\alpha$ working group key paper on DESI year 1 data will be centred around measuring ($\alpha_\parallel$,$\alpha_\perp$), but studies of the Alcock-Paczyński (AP) effect using both the smooth and peak part of the Ly$\alpha$ forest correlations will happen.}

In section \ref{sec:data} we outlined how we compute the Ly$\alpha$ flux transmission catalogue using the full resolution of the DESI spectrograph (0.8$\mathrm{\AA}$). Using the optimal weighting of equation \ref{eq:lya_weight}, by setting $\sigma^2_\mathrm{mod}$=7.5, we reach the same precision on the correlation function as using 2.4$\mathrm{\AA}$ pixels with $\sigma^2_\mathrm{mod}$=3. However, when computing the fits from the latter weighting, we find a $\chi^2$ probability of p=0.13 for the combined fit. Despite this higher fit probability, we prefer to use the full resolution of the DESI spectrograph in this analysis and leave this effect for study in the DESI year 1 analysis. 
Reducing the maximum rest-frame wavelength of the Ly$\alpha$ forest to 1200$\mathrm{\AA}$ from 1205$\mathrm{\AA}$ (chosen to minimise correlation function uncertainties \cite{cper}) increases the fit probability to p=0.18, but we refrain from making this change until there is a further study on mock data sets with greater statistical power. Several sources of contamination like DLAs, metals, and other effects such as quasar radiation are modelled in the same way as in {\dMdB} and require an in-depth study for DESI. 

\subsection{Comparison with eBOSS DR16}
\begin{figure}
\hspace*{-2cm} 
    \centering  
    \includegraphics[scale=0.5]{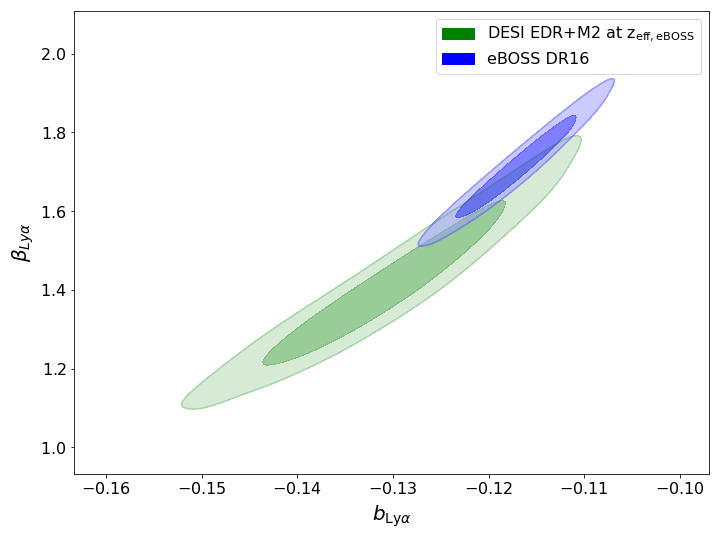}
    \caption{68\% and 95\% credible regions on bias and redshift-space distortions from the auto- and cross-correlation of the Ly$\alpha$ forest for DESI EDR+M2 data at the effective redshift of eBOSS DR16 (green) and eBOSS DR16 data. We performed fits on both datasets with the same model (presented in section \ref{sec:modelling}). To evolve the value of the DESI EDR+M2 bias with redshift we use the relation $b_\mathrm{Ly\alpha} \propto (1+z)^{\gamma-1}$ \cite{McDonald2006} where $\gamma=$2.9\cite{dmdb2020}.}
    \label{fig:bias_zshift}
\end{figure}

To compare with {\dMdB}, we fit their measured correlation functions with the settings described in this paper. Except for the instrumental systematics effect, every other part of the model is the same. In figure \ref{fig:bias_zshift} we present the marginalized 2D posterior of $\beta_\mathrm{Ly\alpha}$ and $b_\mathrm{Ly\alpha}$. Note that the effective redshift of the eBOSS combined correlation is z=2.334, and thus we use the Ly$\alpha$ bias-redshift relation in equation \ref{eq:biasshift} to shift our bias measurement to that effective redshift.

We summarise the differences in the mean value, 68\% confidence intervals, and the consistency between $b_\mathrm{Ly\alpha}$, $\beta_\mathrm{Ly\alpha}$ and a selection of parameters between DESI EDR+M2 and the eBOSS DR16 fits in table \ref{table:compare_params}. In the case of $b_\mathrm{q}$, we also shift the DESI EDR+M2 measurement to the effective redshift of the DR16 data set, this time using the relation $b_\mathrm{q}(\mathrm{z})\propto(1+\mathrm{z})^\gamma$ where $\gamma \sim 1.44$ \cite{dmdb2019}. For the other two parameters in table \ref{table:compare_params}, we are only interested in the mean value and the {relative values of the uncertainties between surveys}. The $\sigma_\mathrm{v}$ constraint tells us that the DESI pipeline more precisely estimates quasar redshifts than was done by eBOSS DR16. 

The final tracers that we constrain in our model are HCDs and metals. We measure the bias of HCDs to be $b_\mathrm{HCD}=$-0.039$\pm$0.009 and $b_\mathrm{HCD}=$-0.053$\pm$0.0043 in early DESI and eBOSS (data) respectively. A smaller value of $b_\mathrm{HCD}$ indicates weaker contamination by DLAs in our data, from which we conclude that the DESI pipeline \cite{jzhou} masks a higher proportion of DLAs, than the {\dMdB} pipeline. The metals are all detected with high confidence, each with a non-zero bias to a minimum of 2.25$\sigma$ except SiII(1193), and are in good agreement with {\dMdB}. The effect of metal correlations is visible in figure \ref{fig:auto_wedge_high}. 

{The systematic error in the quasar redshift estimations is parameterised by $\Delta{\mathrm{r}}_{\parallel,\mathrm{q}}$, which is measured through an asymmetry in the cross-correlation as described in the previous section. We report a value of $\Delta{\mathrm{r}}_{\parallel,\mathrm{q}}$ = -2.21$\pm$0.18$h^{-1}$Mpc, a clear detection of systematic redshift errors.  This offset could be related to an inconsistency in the treatment of the mean flux transmission $\overline{F}(z)$ in the quasar templates used by \texttt{Redrock} at $z\gtrsim$2. In order to avoid this type of bias, {\dMdB} used a redshift estimator which did not use wavelengths around the \lya emission line or lower. They measured a value of $\Delta{\mathrm{r}}_{\parallel,\mathrm{q}}$ = 0.10$\pm$0.11$h^{-1}$Mpc, consistent with zero. The impact of redshift errors in DESI quasars is studied in more detail in \cite{abault}, and mitigation strategies are currently being considered in DESI.}

At $r_\perp=r_\parallel=0$, the spurious correlation produced by instrumental effects is A$_\mathrm{inst}=$2.4$^{+0.3}_{-0.5}\times10^4$. In {\dMdB}, they used a Gaussian instrumental effects model, in which the spurious correlation at $r_\perp=r_\parallel=0$ is A$_{\rm sky}$/$\sigma_{\rm sky}\sqrt{2\pi}=1\times10^4$. We expect to see discrepancies in amplitudes here, given how distinct the two instruments are (section \ref{sec:modelling}). To summarise the differences between early DESI and eBOSS DR16 data discussed above, we include a 13-parameter corner plot for both data sets in Appendix A, figure \ref{fig:full_corn}.

\begin{table}
\centering
\begin{tabular}{lllll}
\hline
\hline
                 Parameter & DESI EDR+M2 &eBOSS DR16& $\sigma_\mathrm{DESI}$/$\sigma_\mathrm{eBOSS}$ & $\Delta\sigma$\\
\hline
\hline
    $\mathit{b}_{\mathrm{Ly}\alpha}$ (z=z$_\mathrm{eff,DR16}$)  &  -0.131$\pm$0.009 &  -0.117$\pm$0.004 & 2.25 & 1.39\\
    $\beta_{\mathrm{Ly}\alpha}$ &  1.41$^{+0.12}_{-0.15}$    & 1.71$\pm$0.09  & 1.5 & 1.88\\
	$\mathit{b}_{\mathrm{q}}$  (z=z$_\mathrm{eff,DR16}$)  & 3.36$\pm$0.16   &  3.70$\pm$0.10&1.6& 1.8 \\
 $\sigma_{\rm v} [h^{-1}\mathrm{Mpc}]$ & 5.2$\pm$0.5&  6.77$\pm$0.29&1.72 & - \\ 
 A$_{\mathrm{BAO}}$ &  1.21$\pm$0.32 & 1.16$\pm$0.16 & 2 & - \\
\hline
\hline
 
\end{tabular}
\caption{Table highlighting the main sampled posteriors of the auto+cross correlations for DESI EDR+M2 and eBOSS DR16 data, their relative uncertainties (68\% credible region), and their consistency with each other, defined as $\Delta\sigma$=$|\mu_{\rm eBOSS}-\mu_{\rm DESI}|$/$\sqrt{\sigma_{\rm eBOSS}^2 + \sigma_{\rm DESI}^2}$, where $\mu$ is the mean parameter value. We use the same model (outlined in section \ref{sec:modelling}) for the fits in both cases, except for instrumental systematics contribution that is slightly different because of variations in the eBOSS and DESI instruments \cite{Satya}.}
\label{table:compare_params}
\end{table}

%% file: conclusion.tex
\section{Conclusion}\label{sec:conclusion}
We present the first study of the Ly$\alpha$ auto-correlation and its cross-correlation with quasars from Dark Energy Spectroscopic Instrument (DESI) data. Our data sample, EDR+M2, consists of 318\,691 quasar target spectra from the DESI survey validation \cite{SVoverview} phase and the first two months of the DESI main survey, resulting in 88\,509 Ly$\alpha$ forests with redshift $z>$2.

{We use the catalogue of \lya fluctuations from \cite{cper}, which made several adaptations to previous analysis pipelines \cite{dmdb2020}, leading to a 20\% and 10\% improvement in the auto- and cross-correlations.} Our 2-point correlation functions are estimated following the eBOSS DR16 analysis \cite{dmdb2020}, with which we report high consistency and an average of 1.7x larger uncertainties across all bins in co-moving separation ($r_\parallel$,$r_\perp$). We do a combined auto+cross correlation fit with our 13 free parameter model, based on linear perturbation theory, and make constraints on key parameters like the \lya forest bias, quasar bias, and the amplitude of the BAO peak. We detect Baryon Acoustic Oscillations to 3.8$\sigma$ confidence, showing the constraining power of DESI data at an early survey stage. The model used is again similar to that in \cite{dmdb2020}, except the instrumental systematics model \cite{Satya} that is modified to account for the differences between both instruments.

At this stage of the DESI survey, we have highlighted the quality of our data and validated existing analysis methods of 3D correlations using the Ly$\alpha$ forest. With relatively low statistical power, we do not report constraints on the scale parameters of BAO and instead leave that to the DESI year 1 analysis. This future work will require an in-depth study of systematics and contaminants like DLAs and metals before performing a proper cosmological analysis of BAO.

%% file: corner_plots.tex
\section{Complete fit results} \label{app:corner_plots}
\begin{figure}
\hspace*{-2cm} 
    \centering  
    \includegraphics[width=1.25\textwidth]{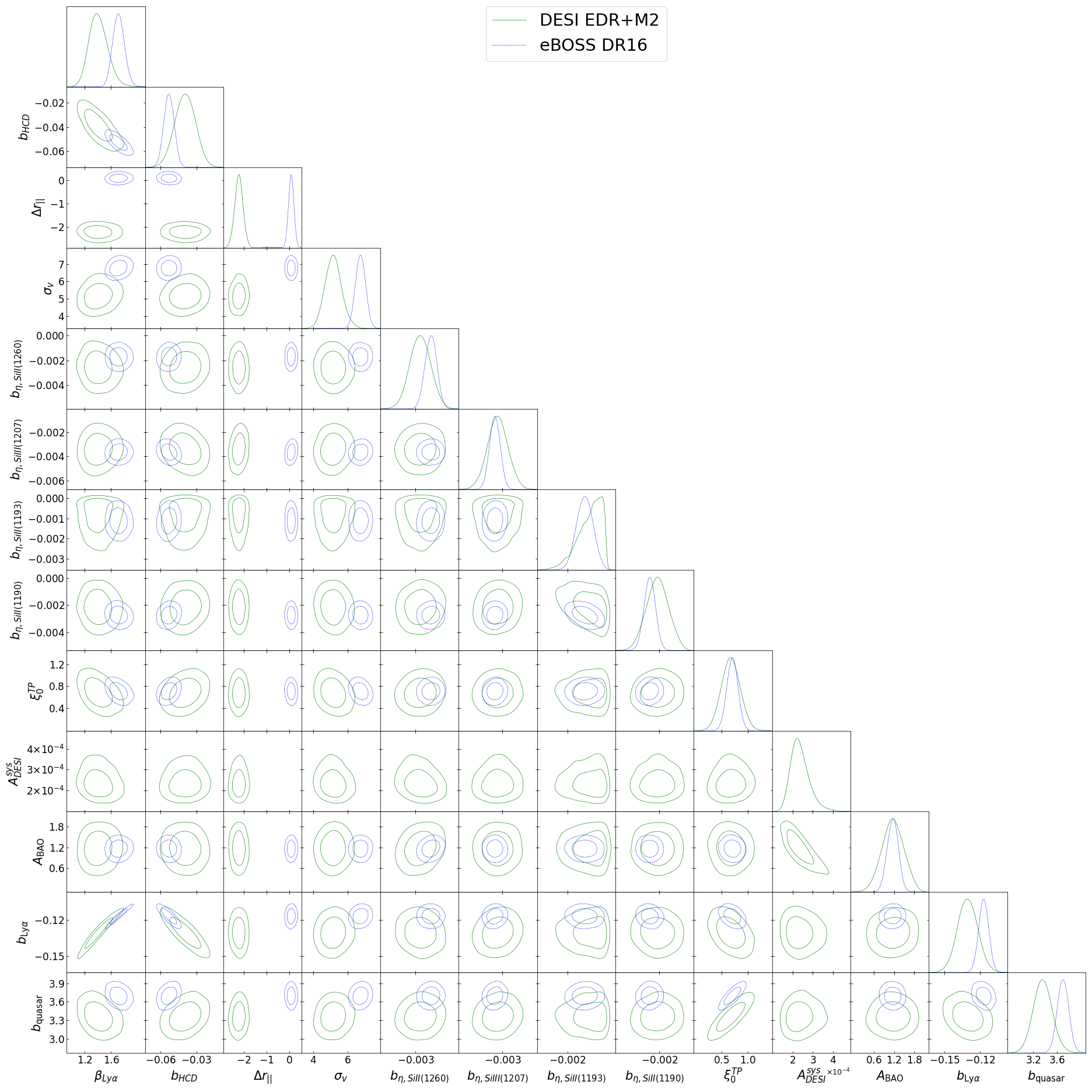}
    \caption{Baseline 13 free parameter fit on DESI EDR and eBOSS DR16 auto+cross correlations. We model the instrumental systematics effects of eBOSS with a slightly different model with two parameters (as opposed to just A$^\mathrm{sky}_\mathrm{DESI}$ for DESI), which we have not included here. For $\mathit{b}_{\mathrm{Ly}\alpha}$ and $b_{\mathrm{q}}$ we present posteriors for both surveys at the effective redshift of the eBOSS DR16 data set.}
    \label{fig:full_corn}
\end{figure}
Here we show the triangle plot of the full posterior distribution - made using the Polychord nested sampler \cite{polychord} and the plotting tool \texttt{Getdist} \cite{getdist} - of all 13 parameters in our model of the combined auto+cross correlations, for both the DESI EDR+M2 and eBOSS DR16 data set. We measure both $b_{\mathrm{Ly}\alpha}$ and $b_{\mathrm{q}}$ at the effective redshift of eBOSS DR16 by using the redshift evolution of each parameter on the EDR+M2 result. For the Ly$\alpha$ and quasars this is $b\propto(1+z)^{\gamma}$, where $\gamma_{\mathrm{Ly\alpha}}=$2.9 \cite{McDonald2006,dmdb2020} and $\gamma_{\mathrm{quasar}}=$1.44 \cite{dmdb2019}.

We can see several parameters in the model which are correlated. The bias and RSD parameters of the Ly$\alpha$ forest are strongly correlated with each other and with the bias factor of HCDs. The correlations that we show between the quasar bias and the bias of HCDs, and the quasar radiation strength are why we do not fit the cross-correlation independently of the auto-correlation, as mentioned in section \ref{sec:fitter}. 

Finally, there is an apparent correlation between the instrumental systematic effects parameter ${\rm A}_{\mathrm{inst}}$ and the BAO peak amplitude A$_{\mathrm{BAO}}$. This is somewhat unexpected because the instrumental systematics model should only contribute at small scales ($r_\parallel=$0,$r_\perp<$40$h^{-1}$Mpc) and needs to be studied further for future analyses of DESI data.

For each parameter, we see the EDR+M2 result is generally within 1-2$\sigma$ of the DR16 results, except $\Delta{{r}}_{\parallel,\mathrm{q}}$, which has a much larger absolute value for our data set. As mentioned in section \ref{sec:fitter}, this is likely because of systematic redshift errors that arise from poorly performing quasar templates. Improvements in the quasar templates used for redshift estimation will be made \cite{qso_templates} for DESI year 1 to reduce this effect, but for now, it is a well-constrained nuisance parameter in our model. Finally, we see that for EDR+M2, the posterior of SiII(1193) is not Gaussian (it is hitting the upper prior bound of 0) because of difficulties constraining the parameter from our data. From the DR16 fits, we can also see that this is one of the weaker detected metals (although it is still non-zero with 3$\sigma$ confidence). Thus we expect that, with lower statistical power, it is difficult to detect. DESI year 1 data will present an exciting opportunity to make the most precise constraints on the presence of metals in the Ly$\alpha$ forest, including lines that are not included in our model here but were in the past (CIV(eff)\cite{dmdb2020}).

%% file: acknowledgments.tex
CG is partially supported by the Spanish Ministry of Science and Innovation (MICINN) under grants PGC-2018-094773-B-C31 and SEV-2016-0588. AC acknowledges support provided by NASA through the NASA Hubble Fellowship grant HST-HF2-51526.001-A awarded by the Space Telescope Science Institute, which is operated by the Association of Universities for Research in Astronomy, Incorporated, under NASA contract NAS5-26555. Both JCM and AFR from the European Union’s Horizon Europe research and innovation programme (COSMO-LYA, grant agreement 101044612). AFR acknowledges financial support from the Spanish Ministry of Science and Innovation under the Ramon y Cajal program (RYC-2018-025210) and the PGC2021-123012NB-C41 project. IFAE is partially funded by the CERCA program of the Generalitat de Catalunya. AXGM acknowledges support from Dirección de Apoyo a la Investigación y al Posgrado, Universidad de Guanajuato, research Grant No. 179/2023 and CONACyT México under Grants No. 286897 and A1-S-17899.

This material is based upon work supported by the U.S. Department of Energy (DOE), Office of Science, Office of High-Energy Physics, under Contract No. DE–AC02–05CH11231, and by the National Energy Research Scientific Computing Center, a DOE Office of Science User Facility under the same contract. Additional support for DESI was provided by the U.S. National Science Foundation (NSF), Division of Astronomical Sciences under Contract No. AST-0950945 to the NSF’s National Optical-Infrared Astronomy Research Laboratory; the Science and Technology Facilities Council of the United Kingdom; the Gordon and Betty Moore Foundation; the Heising-Simons Foundation; the French Alternative Energies and Atomic Energy Commission (CEA); the National Council of Science and Technology of Mexico (CONACYT); the Ministry of Science and Innovation of Spain (MICINN), and by the DESI Member Institutions: \url{https://www.desi.lbl.gov/collaborating-institutions}. Any opinions, findings, and conclusions or recommendations expressed in this material are those of the author(s) and do not necessarily reflect the views of the U. S. National Science Foundation, the U. S. Department of Energy, or any of the listed funding agencies.

The authors are honored to be permitted to conduct scientific research on Iolkam Du’ag (Kitt Peak), a mountain with particular significance to the Tohono O’odham Nation.

%% file: zenodo.tex
\section*{Data points}
The data points corresponding to each figure in this paper can be accessed in the Zenodo repository at \url{https://doi.org/10.5281/zenodo.8244702}.